\documentclass[aps,prb,reprint]{revtex4-2}
\usepackage{graphicx} 
\usepackage{amsmath} 
\usepackage{bm} 
\usepackage{natbib}
\usepackage{natmove}
\usepackage[colorlinks=true, linkcolor=blue, citecolor=blue, urlcolor=blue, linktoc=page, bookmarks=false, pdfstartview={FitH}, pdfborder={0 0 0.0 [3 3]}]{hyperref} 
\usepackage{color} 
\usepackage{dcolumn} 
\newcolumntype{.}{D{.}{.}{2.1}}
\newcolumntype{-}{D{.}{.}{4.0}}
\usepackage{makecell}
\usepackage{multirow}
\usepackage{cleveref}
\usepackage{easyReview} 
\crefname{figure}{Fig.}{Figures}
\crefname{table}{Table}{Tables}
\renewcommand{\today}{\number\day \space \ifcase \month \or January\or February\or March\or April\or May\or June\or July\or August\or September\or October\or November\or December\fi \space \number\year} 
\def\m1r{\multicolumn{1}{r}}
\begin{document}
\title{Perovskite Oxide Heterojunction for Rashba-Dresselhaus Assisted Antiferromagnetic Spintronics}
\author{Jayita \surname{Chakraborty}}
\email{Jayita.Chakraborty1@gmail.com}
\affiliation{Department of Physics, Indian Institute of Science Education and Research Bhopal, Bhauri, Bhopal 462066, India}
\author{Nirmal \surname{Ganguli}}
\email{NGanguli@iiserb.ac.in}
\affiliation{Department of Physics, Indian Institute of Science Education and Research Bhopal, Bhauri, Bhopal 462066, India}
\begin{abstract}
A major impediment towards realizing technologies based on the emerging principles of antiferromagnetic spintronics is the shortage of suitable materials. In this paper, we propose a design of polar$|$nonpolar heterostructure of perovskite oxides, where a single unit cell of SrIrO$_3$ is sandwiched between a thin film of LaAlO$_3$ and a substrate of SrTiO$_3$. Our calculations within the framework of density functional theory + Hubbard $U$ + spin-orbit coupling reveal a two-dimensional conducting layer with electron and hole pockets at the interface, exhibiting strong anisotropic Rashba-Dresselhaus effect along with noncollinear antiferromagnetism, indicating the possibility to realizing a spin-orbit torque. An insightful physical model for anisotropic Rashba-Dresselhaus effect nicely interprets our results, providing an estimate for the Rashba-Dresselhaus coefficients and illustrating pseudospin orientation. We also observe proximity-induced prominent Rashba-like effect for Ti-$3d$ empty bands. Our results suggest that the heterostructure may possess the essential ingredients for antiferromagnetic spintronics, deserving experimental verification.
\end{abstract}
\maketitle

\section{\label{sec:intro}Introduction}
The use of antiferromagnetic materials for spintronic applications has recently been identified as one of the most promising paths for future technology development \cite{BaltzRMP18, ZeleznyNP18, GomonayNP18, JungwirthNN16}. Some of the advantages of antiferromagnetic materials over its conventional ferromagnetic counterparts include the robustness against perturbation due to magnetic fields, the absence of stray fields, and ultrafast spin dynamics in the former. Besides potentially facilitating ultrafast computational operations at a low energy cost, antiferromagnetic spintronic materials are promising for non-volatile memory devices with a terahertz electrical writing speed \cite{OlejnikSA18}. Spin-orbit torque due to strong Rashba-like spin-orbit interaction may help manipulate the magnetic domains and antiferromagnetic spin textures \cite{BaltzRMP18}. Therefore, realizing a substantial Rashba-like effect in a two-dimensional conducting layer along with antiferromagnetism is one of the possible routes for further developments of antiferromagnetic spintronics.

Heterostructures of iridates hold a substantial promise in the lookout for suitable materials for realizing antiferromagnetic spintronics. The emerging research on iridates reveals fascinating physical properties, owing to the interplay of spin-orbit interaction, electron-electron correlation, and crystal field splitting $-$ all with comparable strength in these systems \cite{KimPRL08, ModicNC14, ChakrabortyPRB18}. Moreover, heterostructures of iridates and other 5$d$ oxides may lead to a new paradigm where constraining the electron movement can trigger novel properties that are useful for technology \cite{MatsunoPRL15, GroenendijkPRL17, OhuchiNC18, BhandariPRB18}. In a pioneering work, Matsuno {\em et al.} \cite{MatsunoPRL15} have synthesized superlattices of [(SrIrO$_3$)$_m$, SrTiO$_3$] using pulsed laser deposition technique and studied the physical properties. Their results point out that besides being an insulator, Ir magnetic moments order over a long-range in canted antiferromagnetic fashion in SrIrO$_3$ for $m = 1$, suggesting its similarity with the layered Sr$_2$IrO$_4$. The discussion indicates intuitively that an ultra-thin film of SrIrO$_3$ deposited on a non-interfering substrate may mimic some of the physical properties of Sr$_2$IrO$_4$, owing to the interaction between Ir atoms along $c$ direction being restricted. Another combined experimental and theoretical study suggests that the opening of a gap in an ultra-thin film of SrIrO$_3$ requires antiferromagnetic order \cite{GroenendijkPRL17}. Thin films of SrIrO$_3$ may also exhibit spin-orbit torque \cite{EverhardtPRM19, WangAPL19}. However, a two-dimensional conducting system along with robust antiferromagnetism and strong Rashba-like effect $-$ a combination potentially useful for antiferromagnetic spintronics $-$ has not yet been realized.

In this paper, we propose an ingenious way of realizing a two-dimensional conducting layer with strong spin-orbit interaction leading to the Rashba-Dresselhaus effect in an antiferromagnetic background crucial for hosting antiferromagnetic spintronics. We study a heterostructure having one unit cell of SIO sandwiched between a thick TiO$_2$-terminated SrTiO$_3$ (STO) substrate and a thin film of LaO-terminated LaAlO$_3$ (LAO) along the 001 direction. Thus, the heterostructure of polar LAO with non-polar IrO$_2$-terminated SIO$|$STO is expected to develop an $n$-type interface, where the carriers would be confined to IrO$_2$ plane, resulting in a two-dimensional conducting layer at the interface. Since the bandgap (Mott-like gap) of SIO is very small, an estimate of charge transfer similar to ref.~\onlinecite{GanguliPRL14} would suggest that only a few unit-cell thick LAO would be sufficient for transferring $\sim$0.5e per interface unit cell. Our investigation progresses as follows: first we simulate a TiO$_2$-terminated (SrIrO$_3)_1|$(SrTiO$_3)_1$ heterostructure and its $2 \times 2$ supercell in $ab$ plane (henceforth referred to as $2a \times 2b$ cell) to analyze the electronic structure and magnetic properties. Subsequently, a (LaAlO$_3)_{2.5}|$(SrIrO$_3)_1|$(SrTiO$_3)_{4.5}$ heterostructure that ensures $\sim$0.5e per unit cell transfer to the $n$-type interface and its $2a \times 2b$ supercell are simulated. The electronic structure, magnetic properties, and spin-orbit interaction driven Rashba-like effects are thoroughly investigated for these heterostructures with the help of an analytical model. The remainder of the paper is organized as follows: The methodology of our calculations is described in \cref{sec:method}. A detailed discussion of our results is provided in \cref{sec:results,sec:discuss}. Finally, we summarize the work in \cref{sec:conc}.
\section{\label{sec:method}Method}
Although we aim at understanding the physical properties of a heterostructure with ultra-thin films of SIO and LAO deposited on a thick STO substrate, here we simulate superlattices of (SrIrO$_3)_1|$(SrTiO$_3)_1$  and (LaAlO$_3)_{2.5}|$(SrIrO$_3)_1|$(SrTiO$_3)_{3.5}|$(SrIrO$_3)_1$ comprising $n$-type interfaces between LaO and IrO$_2$ planes, subject to periodic boundary condition along every direction in order to capture the essential physics of the heterostructures within affordable computational cost. The choice of the fractional number of unit cells for the constituent materials ensures the transfer of 0.5 electrons per interface unit cell in order to satisfy the preferred oxidation states of the A-site cation and the oxygen anion. $2a \times 2b$ cells are simulated to accommodate tilted IrO$_6$ octahedra and various antiferromagnetic orders. All our calculations of total energy, electronic structure, and magnetic properties of SIO$|$STO and LAO$|$SIO$|$STO heterostructures are carried out using density functional theory within a plane wave basis set along with projector augmented wave (PAW) method \cite{paw}, as implemented in the {\scshape vasp} code \cite{KressePRB93, vasp2}. Generalized gradient approximation (GGA) due to Perdew-Burke-Ernzerhof (PBE) is used for the exchange-correlation functional \cite{pbe}. Some of the key results were verified within local density approximation (LDA) \cite{ldaCA, PerdewPRB81} for consistency. Strong Coulomb correlation in $d$ and $f$ orbitals was described within DFT + Hubbard $U$ (GGA+$U$) method \cite{DudarevPRB98}, with $U - J$ = 2, 1.5, and 10~eV for Ti-$3d$, Ir-$5d$, and La-$4f$ states, respectively. The large $U$ value for La-$4f$ states ensure locating these empty states sufficiently high in energy to avoid any artifacts in the electronic structure arising from them. The Brillouin zone integration is performed within an improved tetrahedron method \cite{BlochlPRB94T} using a $\Gamma$-centered dense $k$-mesh of $15 \times 15 \times 1$ for the three-dimensional bands and Fermi contour calculations, a $9\times9\times1$ mesh for the rest of the calculations involving $2a \times 2b$ cell, and a $19 \times 19 \times 1$ mesh for the calculations involving $1a \times 1b$ cell. The system's total energy was well-converged to a threshold of 10$^{-7}$~eV for the calculations considering spin-orbit interaction. Mimicking a possible epitaxial growth of the thin films of SIO and LAO on a thick substrate of STO, we have constrained the lattice constant in the $ab$ plane to match that of STO, allowing relaxation only along the $c$ direction to minimize the stress. The atomic positions are optimized to minimize the Hellman-Feynman forces on each atom with a tolerance value of 0.01 eV/\AA.
\section{\label{sec:results}Results}
The results obtained from our calculations are discussed below.
\subsection{SrIrO$_3|$SrTiO$_3$}
\begin{figure}
    \centering
    \includegraphics[scale = 0.46]{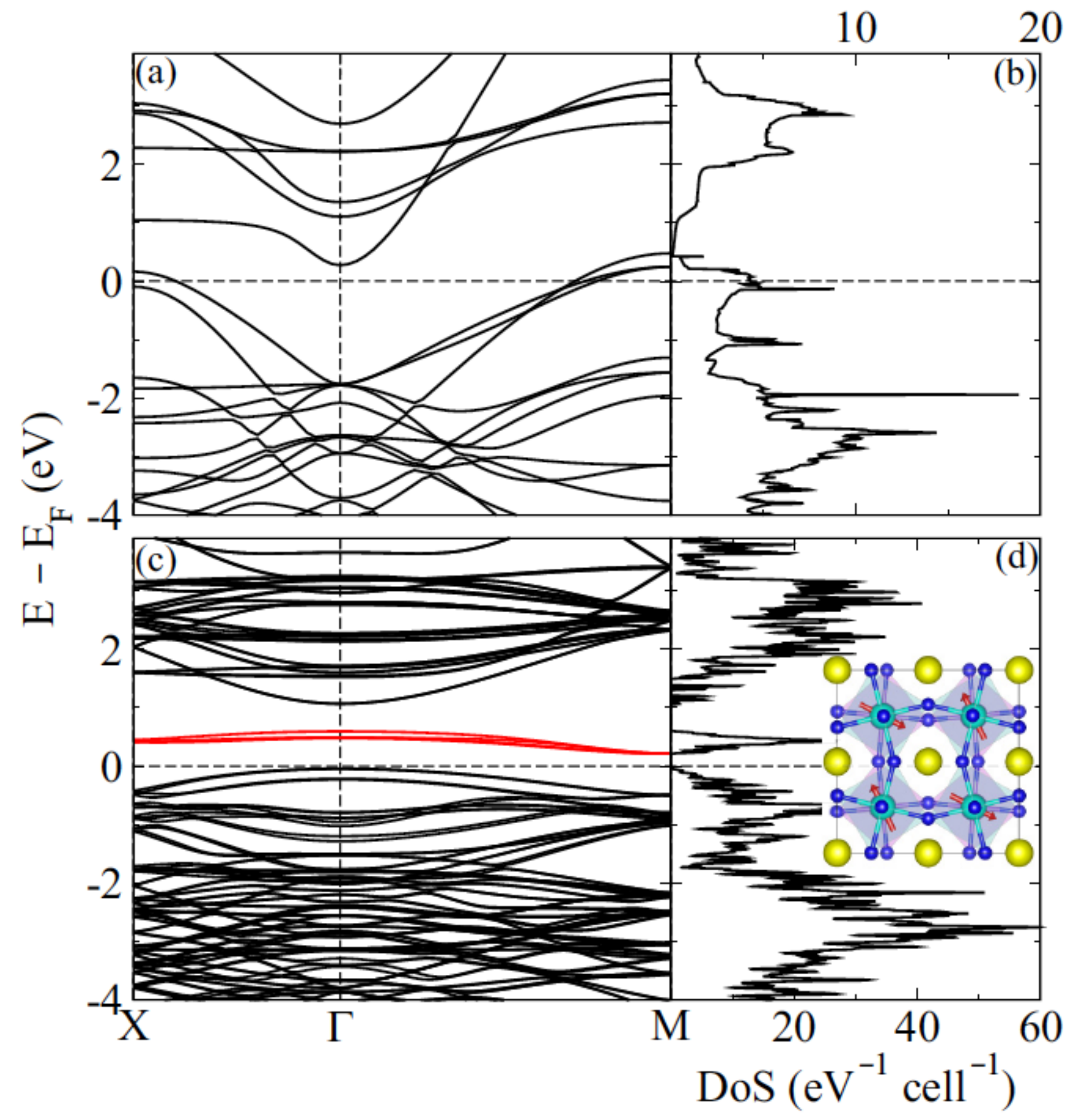}
    \caption{\label{fig:SIOSTOBandDoS}The band structure along the X $\to$ $\Gamma$ $\to$ M direction for SIO$|$STO with $1a \times 1b$ structure (spin-unpolarized) and with $2a \times 2b$ supercell (antiferromagnetic) with spin-orbit interaction are shown in (a) and (c), respectively. (b) and (d) exhibit the density of states corresponding to (a) and (c), respectively. The inset of (d) depicts the tilted IrO$_6$ octahedra and a canted magnetic arrangement.}
\end{figure}
Since SIO$|$STO heterostructure plays the role of a benchmark for our study of LAO$|$SIO$|$STO heterostructure, carefully understanding the electronic structure and magnetic properties of SIO$|$STO is imperative for further progress. Below we discuss the results of our calculations with (SrIrO$_3$)$_1|$(SrTiO$_3$)$_1$ heterostructure. Because each Ir (Ti) atom is surrounded by six O atoms forming an octahedron, Ir-$5d$ (Ti-$3d$) spin-degenerate states split into a lower energy three-fold degenerate $t_{2g}$ manifold and a higher energy two-fold degenerate $e_g$ manifold. Further, strong spin-orbit interaction in an Ir$^{4+}$ ion splits the $t_{2g}$ manifold into a lower-energy completely filled $J_\text{eff} = 3/2$ quartet and a higher-energy half-filled $J_\text{eff} = 1/2$ doublet. Our results for the calculations with $1a \times 1b$ structure reveals the conducting nature of the heterostructure, owing to Ir-$5d$ bands crossing the Fermi level. The band dispersion along $X \to \Gamma \to M$ direction (the Brillouin zone geometry has been discussed in \cref{sec:BZ}) and the density of states (DoS) are displayed in \cref{fig:SIOSTOBandDoS}(a) and \cref{fig:SIOSTOBandDoS}(b), respectively. We note that Ir-$5d$ states, hybridized with O-$2p$ states, constitutes the states near the Fermi level ($E_F$). Ti-$3d$ states in Ti$^{4+}$ ions are completely empty and lie above $E_F$. The heterostructure's metallic character originates from the Ir-$5d-$O-$2p$ hybridized states, suggesting that the conduction electrons are confined to the single layer of SrIrO$_3$, leading to a two-dimensional conducting layer. To allow antiferromagnetic order or tilted IrO$_6$ octahedra, we have simulated a $2a \times 2b$ supercell. In practice, a heterostructure with an ultra-thin film of SIO grown on a thick substrate of STO would adapt to STO's lattice constant in the $ab$ plane. On the other hand, the lattice constant along the $c$-direction is unconstrained and governed by the Poisson's ratio of the respective materials. When deposited on the STO substrate, SIO undergoes an in-plane compressive strain. Inset of \cref{fig:SIOSTOBandDoS}(d) shows the optimized structure with tilted IrO$_6$ octahedra having a $a^0b^0c^-$ tilt pattern with reference to the nearby STO layer \cite{GlazerACB72} and an angle of 14.1$^\circ$ about the $c$-axis, in agreement with experiments \cite{BiswasJAP14, GruenewaldJMR14, FanJPCS15}. Our calculations for $2a \times 2b$ supercell, including spin-orbit interaction, suggest that a ferromagnetic order is not stable. We realize a weak canted antiferromagnetic arrangement, as depicted with arrows in the inset of \cref{fig:SIOSTOBandDoS}(d), with projected spin and orbital magnetic moments of 0.12 and 0.27~$\mu_B$, respectively, at Ir site. The band structure (\cref{fig:SIOSTOBandDoS}(c)) and the density of states (\cref{fig:SIOSTOBandDoS}(d)) corresponding to the antiferromagnetic phase clearly shows a small energy gap of 0.16 eV near the Fermi level. Such a bandgap opening, driven by antiferromagnetic order, with a moderate value of $U - J = 2$~eV within GGA+$U$ method suggests a Mott- or Slater-like insulating state of the monolayer of SrIrO$_3$, owing to the absence of interaction between Ir atoms along the $c$-direction similar to Sr$_2$IrO$_4$ \cite{WatanabePRB14, SchutzPRL17}. Our results do not reveal any Rashba-like splitting of the bands, consistent with the experimental report of temperature-dependent weak Rashba-like effect for SIO$|$STO heterostructure (see \cref{sec:Rashba} for further details) \cite{ZhangJPSJ14}. An isolated narrow (0.37~eV wide) $J_\text{eff} = 1/2$ band appears right above the Fermi level that would lead to interesting physics upon charge transfer.
\subsection{LaAlO$_3|$SrIrO$_3|$SrTiO$_3$}
Similar to the polar-nonpolar oxide heterostructure of LAO$|$STO, deposition of a thin film of the polar oxide LAO on the one unit cell thick film of IrO$_2$-terminated non-polar SIO grown on non-polar STO substrate is expected to result in an $n$-type interface with the two-dimensional conducting layer being confined at the IrO$_2$ plane.

\subsubsection{Tilted IrO$_6$ Octahedra}
We optimized the atomic positions and the lattice constant along $c$-direction for the simulated LAO$|$SIO$|$STO heterostructure, constraining the lattice constant in the $ab$ plane to that of STO. An $a^0b^-c^-$ tilt pattern emerges from our calculations, with tilt angles of $5.4^\circ$ and $13.4^\circ$ about $b$ and $c$ axes, respectively (see \cref{fig:ChargeDensity}(a) and \cref{fig:ChargeDensity}(b) inset) that lowers the total energy of the system by 0.45~eV per Ir atom when compared with a heterostructure without octahedral tilts. The enhanced tilts can further confine the electrons in the local orbitals, reinforcing magnetism \cite{GanguliPRL14}. We note that in the absence of experiments, the possible thickness of LAO film that can be grown on the ultra-thin film of SIO is not known. At this stage, an estimate of the charge transfer amount to the interface would help understand the proposed system's physical properties.
\subsubsection{Charge Transfer}
\begin{figure}
    \centering
    \includegraphics[scale = 0.44]{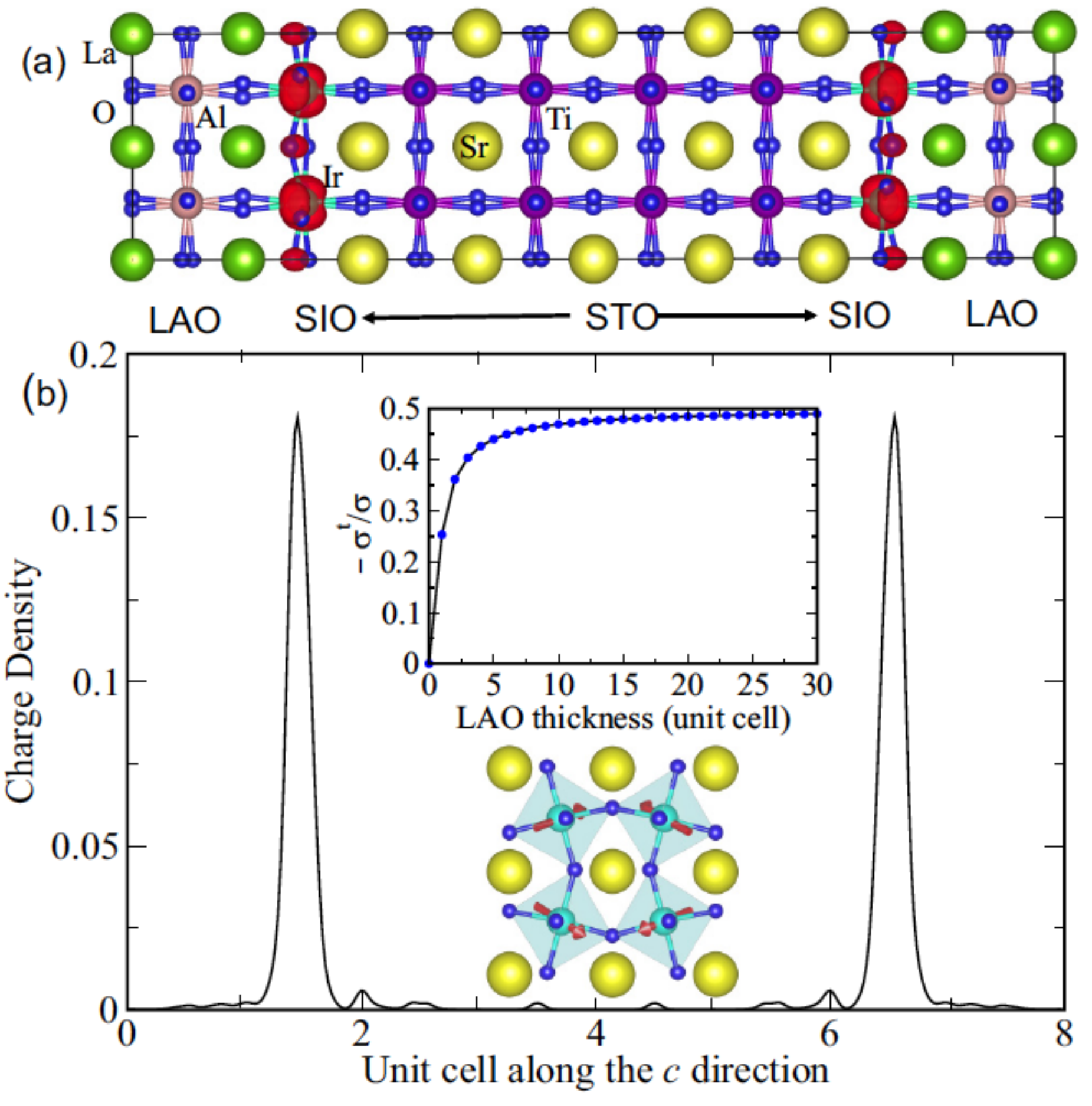}
    \caption{\label{fig:ChargeDensity}Subfigure (a) depicts the transferred charge density isosurfaces for the LAO$|$SIO$|$STO heterostructure, while (b) shows the transferred charge density summed over $ab$ plane varying with the unit cells along the $c$-direction. An electrostatic estimate of charge transfer as a function of LaAlO$_3$ thickness and an illustration of the tilted IrO$_6$ octahedra marking the magnetic moment vectors are given in the insets of (b).}
\end{figure}
Our estimate of charge transfer is based on electrostatics, as described in ref.~\onlinecite{GanguliPRL14}. Using appropriate values for the parameters \footnote{We have used $\epsilon^\text{LAO}/\epsilon_0 = 24$, $\epsilon_0$ being the permittivity of free space, $\varepsilon_g + \Delta = 0.35$~eV, $K^\text{IF} = 0.5$~F$^{-1}$}, we obtained the ratio of the required amount of transferred surface charge density $\sigma^t$ to the surface charge density of a positively charged plane $\sigma$ for avoiding polar catastrophe, as shown in the inset of \cref{fig:ChargeDensity}(b). Because of the small Mott-like gap near the Fermi level and the small valence band offset between SrIrO$_3$ and LaAlO$_3$, the required amount of transferred charge rapidly approaches its saturation value of 0.5e$^-$ per interface unit cell with increasing thickness of LAO, reaching 0.4e$^-$ for as thin as three unit-cell of LAO. Hence we justify representing the arrangement of the thin films described above by simulating a LAO$|$SIO$|$STO heterostructure (see \cref{fig:ChargeDensity}(a)) within periodic boundary condition along $c$-direction that ensures the transfer of 0.5e$^-$ per interface unit cell.

The charge density isosurface plot for the transferred charge displayed in \cref{fig:ChargeDensity}(a) finds the transferred charge populating the hybridized Ir-$5d-$O-$2p$ states in the IrO$_2$ plane. Further, we have calculated the transferred charge density $\rho(z)$ within the appropriate energy range, summing over the FFT grid points along $x$ and $y$ directions. The result, shown in \cref{fig:ChargeDensity}(b), confirms that the transferred charge is confined only to the IrO$_2$ planes, corroborating the impression gathered from the isosurface plot (\cref{fig:ChargeDensity}(a)).
\subsubsection{Magnetism}
We note that partially filling the narrow $J_\text{eff} = 1/2$ bands of Ir right above the Fermi level in \cref{fig:SIOSTOBandDoS}(c)  with the electrons transferred due to deposition of a polar oxide film in the presence of heavily tilted IrO$_6$ octahedra is expected to lead to interesting magnetic properties.
\begin{table}
\begin{center}
\caption{\label{tab:EnergyMoment} The spin (orbital) magnetic moment projected onto the Ir ions and the relative energies $\Delta E$ in meV for different magnetic configurations of LAO$|$SIO$|$STO: canted ferromagnetic (FM), canted checkerboard antiferromagnetic (CB-AFM), and canted striped antiferromagnetic (ST-AFM) are tabulated here.}
\begin{ruledtabular}
\begin{tabular}{ccccc}
Configuration & \multicolumn{2}{c}{$U - J = 1.5$~eV} & \multicolumn{2}{c}{$U - J = 2.0$~eV}\\
\cline{2-3} \cline{4-5} 
& Ir moment ($\mu_B$) &  $\Delta E$ &Ir moment ($\mu_B$) & $\Delta E$  \\
\hline
FM & 0.2  (0.21)  & 0.85  & 0.21 (0.23) & 0.68 \\
CB-AFM & 0.2 (0.24) & 0.00 & 0.21 (0.25) & 0.00 \\
ST-AFM & 0.2 (0.24) & 0.53 & 0.21 (0.25) & 0.50 \\
\end{tabular}
\end{ruledtabular}
\end{center}
\end{table}
Our calculations suggest that the Ir ions' magnetic moments are canted along the axes of the tilted IrO$_6$ octahedra. We have simulated three qualitatively different magnetic arrangements: (a) the canted ferromagnetic (FM) arrangement, where the canted magnetic moments point roughly along the same direction, (b) the canted checkerboard antiferromagnetic (CB-AFM) arrangement, where the second nearest magnetic moments are in almost opposite directions, and (c) the canted striped antiferromagnetic (ST-AFM) arrangement, where nearly opposite magnetic moments form stripes. Our results for the projected magnetic moments and relative energies for different magnetic arrangements are tabulated in \cref{tab:EnergyMoment}. The results indicate that the canted CB-AFM is the most favored in terms of energy among the magnetic arrangements considered here. In addition to a substantial projected spin moment of $0.2~\mu_B$, we observe a remarkably significant orbital moment of $0.24~\mu_B$ projected onto the Ir ions, which may be attributed to Ir's strong spin-orbit interaction. We have also repeated the calculations with $U - J = 2$~eV for Ir-$5d$ states and found qualitatively similar results with little increment in the projected magnetic moment (see \cref{tab:EnergyMoment}). The lowest energy canted CB-AFM arrangement is depicted in the inset of \cref{fig:ChargeDensity}(b) with arrows for the projected magnetic moment vectors. To assess the reliability of our results, we repeated the calculations within local spin density approximation + Hubbard $U$ method with a modest value of $U - J = 1.5$~eV for Ir-$5d$ states, considering spin-orbit coupling (LSDA+$U$+SOC). The results reconfirmed the canted CB-AFM as the lowest energy state magnetic arrangement with $0.17~\mu_B$ and $0.23~\mu_B$ of projected spin and orbital moments, respectively, at Ir sites. The agreement of the results obtained within GGA+$U$+SOC and LSDA+$U$+SOC point to the robustness of our predictions for the magnetic properties of the heterostructure. The robust antiferromagnetism with substantial orbital moment makes the heterostructure promising for technologies based on antiferromagnetic spintronics.
\subsubsection{Rashba-like Effect}
\begin{figure*}
    \centering
    \includegraphics[scale = 0.70]{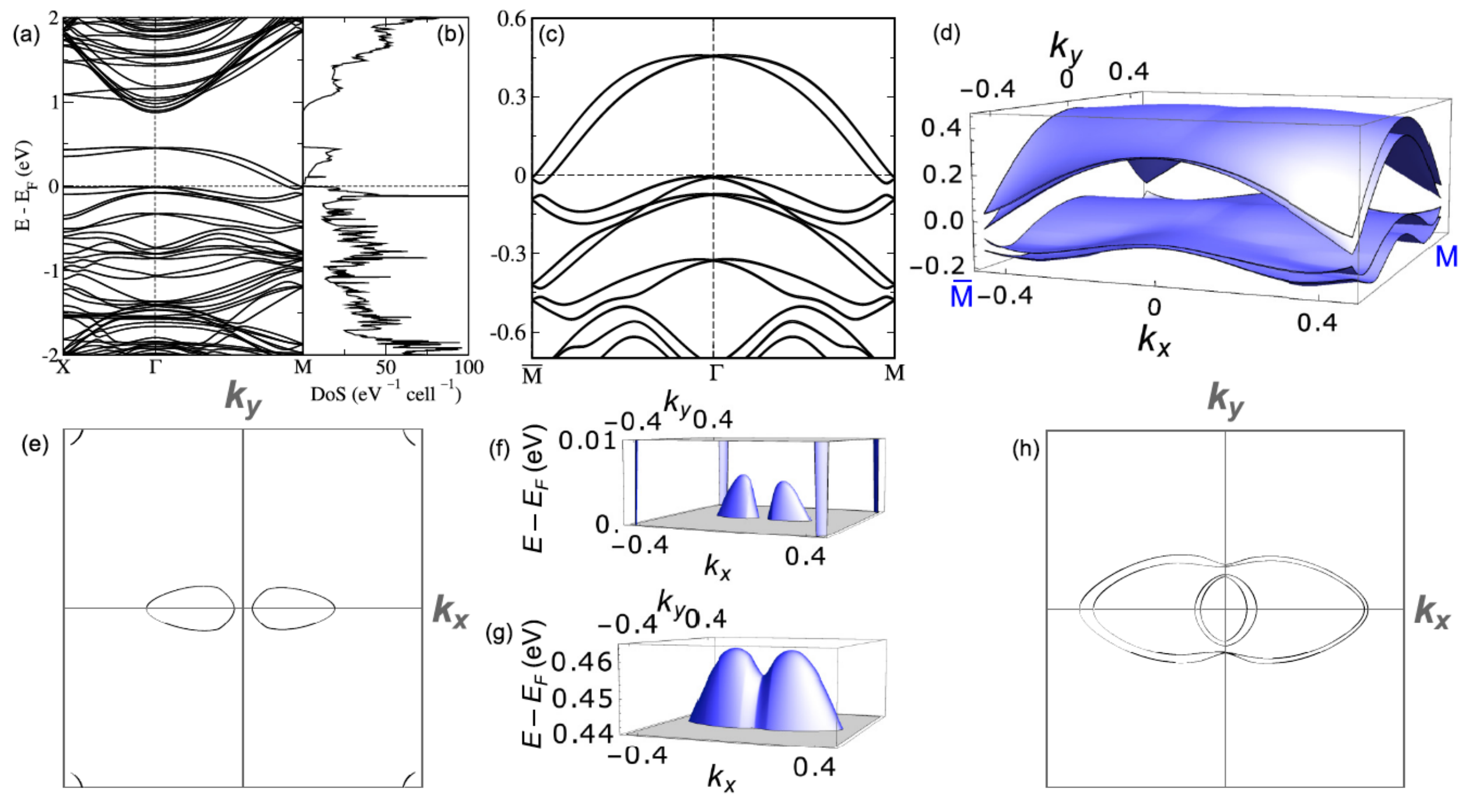}
    \caption{\label{fig:Rashba-Ir}The band dispersion of LAO$|$SIO$|$STO in the canted CB-AFM configuration along the high-symmetry directions $X \to \Gamma \to M$ and the density of states are shown in (a) and (b), respectively. (c) shows the predominantly Ir-$5d$ bands near the Fermi level along $\bar{M} \to \Gamma \to M$, highlighting the Rashba-like splitting. The bands touching the Fermi level have been plotted in three-dimension (3D) as functions of the crystal momenta $k_x$ and $k_y$, displayed in (d). The Fermi contours feature in (e), while the hole pockets are highlighted in a 3D plot, shown in (f). The top of the bands highlighted in (g) has two peaks, leading to double elliptical constant energy contours on the plane $E - E_F = 0.44$~eV, shown in (h).}
\end{figure*}
Having an insight into the magnetic ground state of LAO$|$SIO$|$STO heterostructure, we focus on the consequences of strong spin-orbit interaction of Ir-$5d$ electrons on the physical properties of the system. The results obtained from our electronic structure calculations for the heterostructure within GGA+$U$+SOC are displayed in \cref{fig:Rashba-Ir}. We find from the band dispersion and DoS shown in \cref{fig:Rashba-Ir}(a) and \cref{fig:Rashba-Ir}(b), respectively, that the narrow (0.37~eV wide) Ir-$5d$ band above the Fermi energy highlighted in \cref{fig:SIOSTOBandDoS}(c) gets broadened to $\sim$0.5~eV upon transfer of 0.5e$^-$ per interface unit cell. The formation of the heterostructure and tilts of IrO$_6$ octahedra leads to broken symmetry that splits the Ir bands near the Fermi level, resulting in a dip in the DoS at the Fermi level. As a consequence of broken inversion symmetry in the presence of a microscopic electric field and strong spin-orbit interaction, we expect to see a Rashba-like effect in the system that would separate the bands with different pseudospins in momentum space. \cref{fig:Rashba-Ir}(c) displays the bands near the Fermi level along the $\bar{M} \to \Gamma \to M$ direction, confirming our expectation of Rashba-like split bands. In order to gain further insight into the electronic structure of these bands, we exploit the two-dimensional nature of the Brillouin zone for the heterostructure and plot the bands as functions of $k_x$ and $k_y$, shown in \cref{fig:Rashba-Ir}(d). Examining \cref{fig:Rashba-Ir}(c) and \cref{fig:Rashba-Ir}(d), we observe the presence of electron pockets near the corners of the Brillouin zone and shallow hole pockets near the center of the Brillouin zone. The Fermi contours displayed in \cref{fig:Rashba-Ir}(e) confirm the location of the electron and hole pockets, while \cref{fig:Rashba-Ir}(f) reveals the shape of the hole pockets in three dimensions. A close look at the top of the bands highlighted in \cref{fig:Rashba-Ir}(g) reveals two peaks, which is not a characteristic of the Rashba effect alone. Further, the constant energy contours on the plane $E - E_F = 0.44$~eV (see \cref{fig:Rashba-Ir}(h)) exhibit double lines traced by nearly degenerate bands from different sublattices, a characteristic of constant energy contours for Rashba-like split bands in an antiferromagnetic system \cite{KrupinNJP09}, with elliptical arcs merging onto each other. These features call for a deeper understanding of the Rashba-like physics in the system.
\subsubsection{Anisotropic Rashba-Dresselhaus Model}
\begin{figure*}
    \centering
    \includegraphics[scale = 0.70]{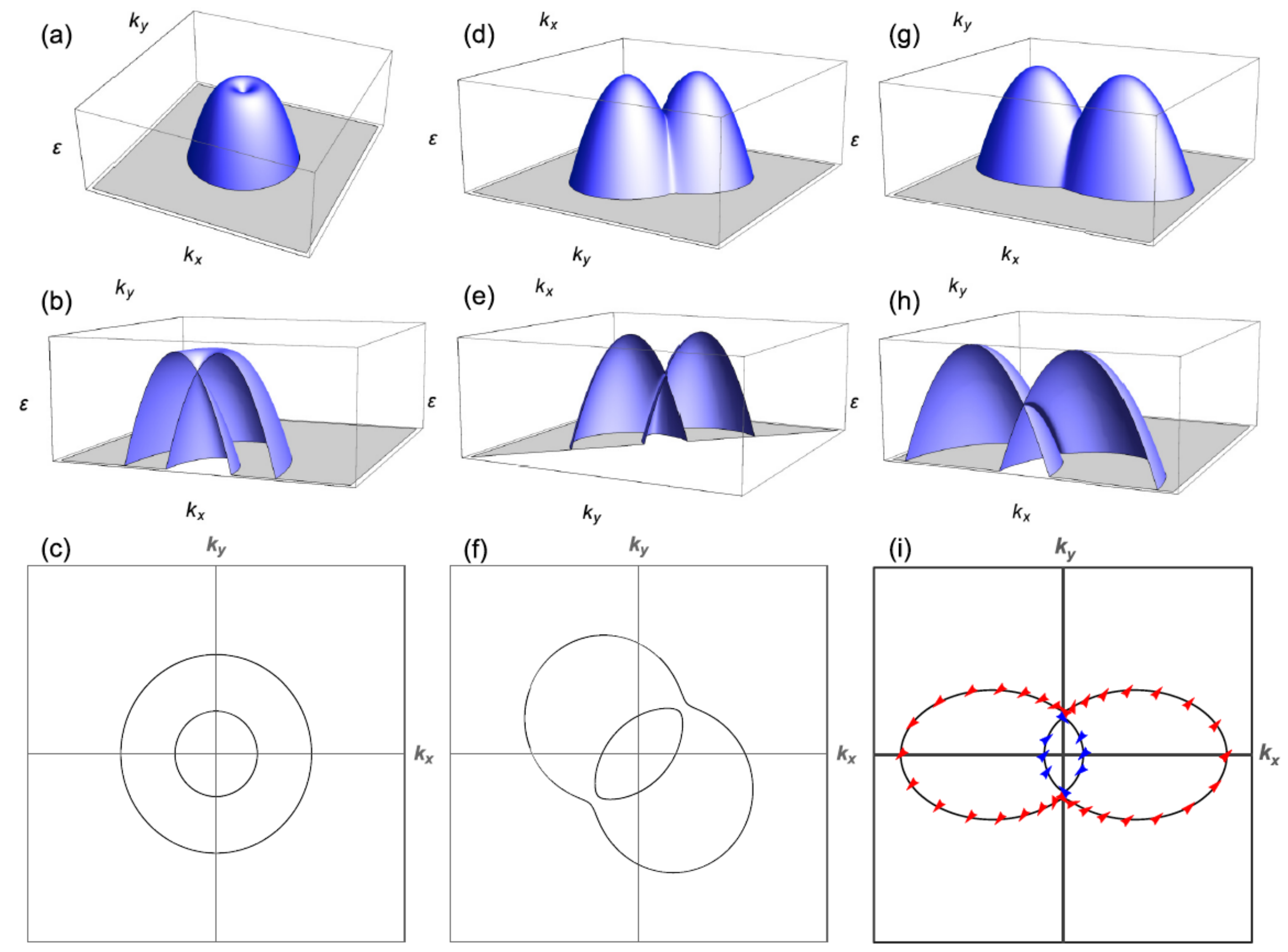}
    \caption{\label{fig:ModelBands}We have shown the band dispersions $\varepsilon(\vec{k})$ and constant energy contours obtained from our model calculations in this figure. Subfigures (a) and (b) show the dispersion for Rashba effect on isotropic bands, while (c) shows the corresponding constant energy contours. Similarly, (d) and (e) shows the band dispersion for a combination of isotropic Rashba and Dresselhaus effects, with the corresponding constant energy contours in (f). Finally, the energy-momentum dispersion from our model for anisotropic Rashba-Dresselhaus effect with rotated Brillouin zone features in (g) and (h), with the corresponding constant energy contours in (i), marking the pseudospin directions with arrows.}
\end{figure*}
\begin{figure}
    \centering
    \includegraphics[scale = 0.6]{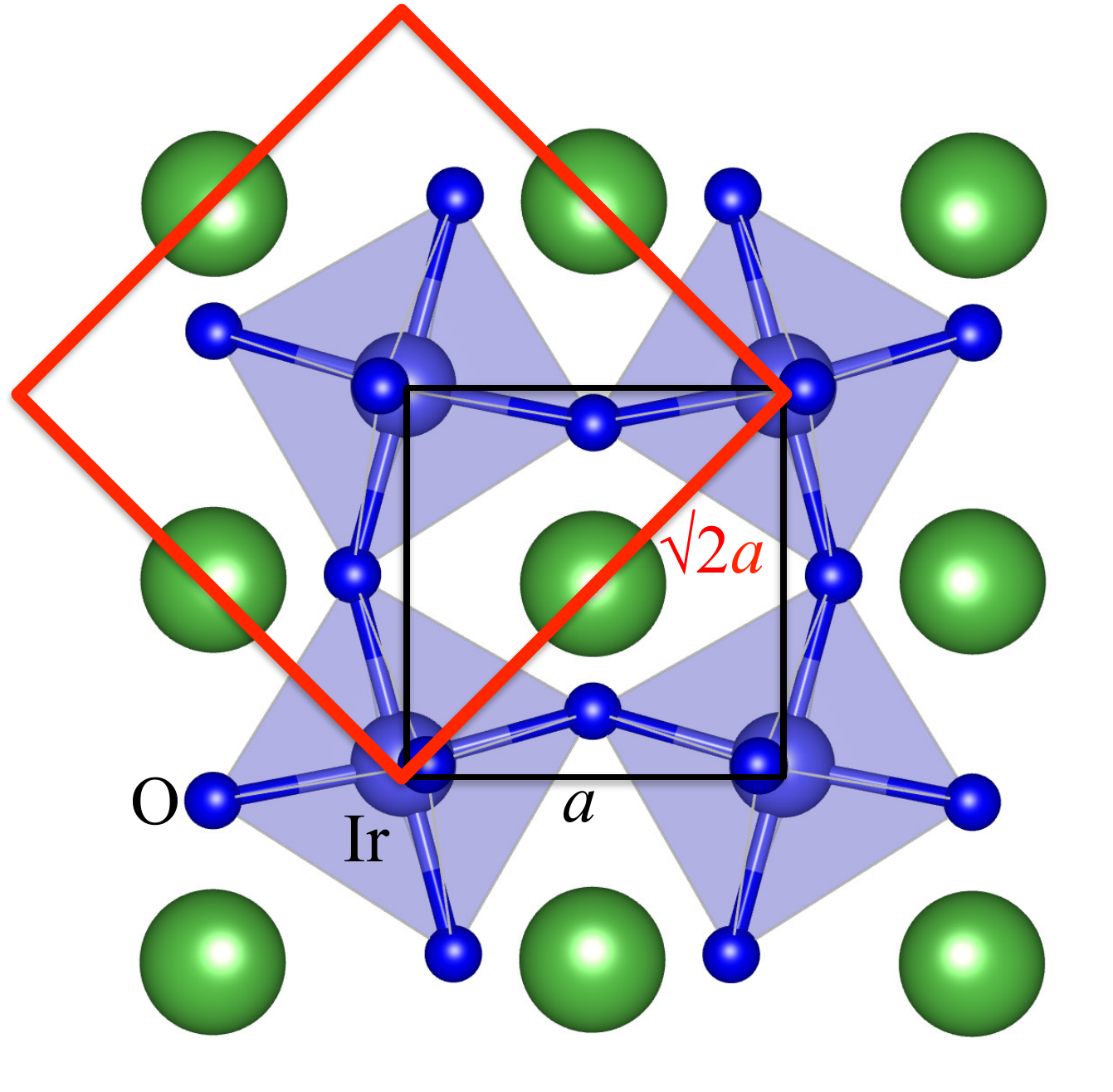}
    \caption{\label{fig:r2xr2}The red square represents the $\sqrt{2} \times \sqrt{2}$ times cell in the $ab$ plane that is required to accommodate the IrO$_6$ octahedra tilted in $a^0b^-c^-$ pattern, while the black square shows the unit cell sufficient to describe the structure without octahedral tilts. They make an angle of $\pi/4$.}
\end{figure}

In order to better understand the unusual Rashba-like features of the system, we develop an analytical model of the anisotropic Rashba-Dresselhaus effect. The influence of an electric field along the $z$-direction on the electrons with strong spin-orbit interaction and structure inversion asymmetry (SIA) may be described by the Rashba Hamiltonian of the form \cite{RashbaSPSS60}
\begin{equation}
	H_R = \alpha ( k_y \sigma_x - k_x \sigma_y), \label{eq:Rashba}
\end{equation}
where $\alpha$ is the Rashba coefficient, $k_x, k_y$ are the components of crystal momentum, and $\sigma_x, \sigma_y$ are the usual Pauli matrices.
Similarly, Dresselhaus Hamiltonian arising due to spin-orbit interaction and bulk inversion asymmetry (BIA) is given as \cite{DresselhausPR55}
\begin{equation}
    H_D = \beta (k_y \sigma_y - k_x \sigma_x), \label{eq:Dresselhaus}
\end{equation}
with $\beta$ being the Dresselhaus coefficient. When treated as a perturbation to a free quasiparticle system with effective mass $m^*$, the Rashba (Dresselhaus) term leads to the eigenvalues
\begin{equation}
	\varepsilon_{R(D)}^{\pm} (\vec{k}) = \frac{k^2}{2m^*} \pm \alpha (\beta) |\vec{k}|,
\end{equation}
adjusting the units such that $\hbar = 1$. The qualitatively similar eigenvalues $\varepsilon_{R(D)}^{\pm} (\vec{k})$ from the Rashba and the Dresselhaus terms plotted as functions of $k_x$ and $k_y$ for $m^* < 0$ are shown in \cref{fig:ModelBands}(a), while \cref{fig:ModelBands}(b) displays a cross-section of the same bands. \cref{fig:ModelBands}(c) displays the circular constant energy contours, obtained by setting
\begin{equation}
	\varepsilon_{R(D)}^{\pm} (\vec{k}) = \frac{k^2}{2m^*} \pm \alpha (\beta) |\vec{k}| = \text{constant}.
\end{equation}
In the presence of both structure and bulk inversion asymmetry in a system, the combined Rashba-Dresselhaus Hamiltonian leads to the eigenvalues
\begin{equation}
	\varepsilon_{RD}^{\pm} (\vec{k}) = \frac{k^2}{2m^*} \pm \sqrt{(\alpha^2 + \beta^2 ) k^2 - 4 \alpha \beta k_x k_y},
\end{equation}
featuring double-hill bands shown in \cref{fig:ModelBands}(d) and a cross-section in \cref{fig:ModelBands}(e). While the double-hill pattern somewhat resembles the band structure obtained from our GGA+$U$+SOC calculations (see \cref{fig:Rashba-Ir}(g)), the constant energy contours in \cref{fig:ModelBands}(f) exhibit significant differences with \cref{fig:Rashba-Ir}(h) in terms of shape and orientation $-$ the former comprises circular arcs with a diagonal orientation, while the latter features elliptical arcs with an axial orientation. We note that the $a^0b^-c^-$ tilt pattern of the IrO$_6$ octahedra with no tilt about the $a$-axis and substantial tilt about the $b$-axis may lead to unequal quasiparticle effective masses along $k_x$ and $k_y$ directions. Further, we must increase the unit cell at least $\sqrt{2} \times \sqrt{2}$ times in the $ab$-plane in order to accommodate the $a^0b^-c^-$ tilt pattern (see \cref{fig:r2xr2}), rotating the reciprocal cell by $\pi/4$ relative to the principal axes of effective mass. The transformed coordinates in the reciprocal space $(K_X, K_Y)$ may be represented in terms of $(k_x, k_y)$ as
\begin{equation}
    \begin{pmatrix}
        K_X \\
        K_Y
    \end{pmatrix} = \frac{1}{\sqrt{2}}
    \begin{pmatrix}
        \cos \frac{\pi}{4} & \sin \frac{\pi}{4} \\
        - \sin \frac{\pi}{4} & \cos \frac{\pi}{4}
    \end{pmatrix}
    \begin{pmatrix}
        k_x \\
        k_y
    \end{pmatrix} = \frac{1}{2}
    \begin{pmatrix}
        k_x + k_y \\
        k_y - k_x
    \end{pmatrix}. \label{eq:Kk}
\end{equation}
Considering the anisotropic effective mass and the transformed reciprocal axes, we can write the Hamiltonian for the anisotropic Rashba-Dresselhaus model as
\begin{equation}
	H_{ARD} = H_0 + \alpha ( K_Y \sigma_x - K_X \sigma_y) + \beta (K_Y \sigma_y - K_X \sigma_x), \label{eq:ARD-H}
\end{equation}
where
\begin{equation}
	H_0 = -\frac{1}{2 m_x^*} \frac{\partial^2}{\partial x^2} - \frac{1}{2 m_y^*} \frac{\partial^2}{\partial y^2} \label{eq:free-H}
\end{equation}
is the kinetic energy term in two dimension, with $m_x^*$ and $m_y^*$ representing the effective masses along $k_x$ and $k_y$, respectively. Treating the Rashba-Dresselhaus terms in \cref{eq:ARD-H} as perturbations to the free-electron-like term $H_0$ (see \cref{eq:free-H}) and using \cref{eq:Kk} for relating $(K_X, K_Y)$ to $(k_x, k_y)$, we evaluate the energy eigenvalues $\varepsilon_{ARD}^{\pm}(\vec{k})$ for the anisotropic Rashba-Dresselhaus system as
\begin{equation}
    \varepsilon_{ARD}^{\pm} (\vec{k}) = \frac{k_x^2}{2 m_x^*} + \frac{k_y^2}{2 m_y^*} \pm \sqrt{2[(\alpha + \beta)^2 k_x^2 + (\alpha - \beta)^2 k_y^2]}. \label{eq:ARD}
\end{equation}
\Cref{fig:ModelBands}(g) and \cref{fig:ModelBands}(h) shows the double-hill bands and their cross-section, respectively.

The constant energy contours may be obtained in polar coordinate system by putting $k_x = k \cos \phi$ and $k_y = k \sin \phi$ where $(k, \phi)$ are the usual polar coordinates, and setting $\varepsilon_{ARD}^{\pm}$ in \cref{eq:ARD} to a constant $\varepsilon_c$, as below:
\begin{align}
    \varepsilon_c &= \frac{k^2 \cos^2 \phi}{2 m_x^*} + \frac{k^2 \sin^2 \phi}{2 m_y^*} \nonumber \\
    &\pm \sqrt{2[(\alpha + \beta)^2 k^2 \cos^2 \phi + (\alpha - \beta)^2 k^2 \sin^2 \phi} \nonumber \\
    &= k^2 \left( \frac{\cos^2 \phi}{2 m_x^*} + \frac{\sin^2 \phi}{2 m_y^*} \right) \pm k \sqrt{2 (\alpha^2 + \beta^2 + 2 \alpha \beta \cos 2 \phi)}
\end{align}
The roots of this quadratic equation give the constant energy contours, as shown in \cref{fig:ModelBands}(i), revealing no qualitative difference with our GGA+$U$+SOC results (see \cref{fig:Rashba-Ir}(h)). The model with the parameter values of $m_x^* = -0.81 m_e$, $m_y^* = -0.29 m_e$, $\alpha = 0.062$~eV\AA, and $\beta = 0.044$~eV\AA\ reasonably emulates our results from GGA+$U$+SOC calculations shown in \cref{fig:Rashba-Ir}(g), suggesting a large Rashba-Dresselhaus effect for the system $-$ nearly two orders of magnitude higher than that of SIO$|$STO heterostructure \cite{ZhangJPSJ14}. Our estimated effective masses belong to similar ballpark range as reported by \citet{MancaPRB18} for the Ir-$5d$ bands in SIO$|$STO heterostructure, although with a different tilt pattern of the IrO$_6$ octahedra. Further, anisotropic transport properties were reported in untrathin SIO$|$STO heterostructure, in agreement with our model of anisotropic effective mass \cite{VanThielACSML20}.

In order to visualize the pseudospin directions in the system, we evaluated the eigenstates of the anisotropic Rashba-Dresselhaus Hamiltonian $H_{ARD}$ as
\begin{equation}
	|\pm \rangle = \frac{1}{\sqrt{2}} (\pm \zeta | \uparrow \rangle + | \downarrow \rangle ),
\end{equation}
where $| \uparrow \rangle$ and $| \downarrow \rangle$ are the eigenstates of the spin projection operator $S_z$, and
\begin{equation}
	\zeta = \frac{(1 - i) \sqrt{\alpha^2 + \beta^2 + 2 \alpha \beta \cos 2 \phi}}{\sqrt{2}[(\alpha + \beta) \cos \phi + i(\alpha - \beta) \sin \phi]}, ~~~ \zeta^* \zeta = 1.
\end{equation}
We may evaluate the expectation values of the spin-projections for $| \pm \rangle$ in the units of $\hbar$ as follows:
\begin{align}
	\left\langle S_z \right\rangle_+ &= \frac{1}{2} \langle + | \sigma_z | + \rangle = \frac{1}{4}(\zeta^* \zeta - 1) = 0 \nonumber \\
	&= \frac{1}{2} \langle - | \sigma_z | - \rangle = \left\langle S_z \right\rangle_-, \label{eq:Sz} \\
	\left\langle S_x \right\rangle_+ &= \frac{1}{2} \langle + | \sigma_x | + \rangle = \frac{1}{4}(\zeta + \zeta^*) \nonumber \\
	&= \frac{1}{2 \sqrt{2}} \frac{(\alpha + \beta) \cos \phi - (\alpha - \beta) \sin \phi}{\sqrt{\alpha^2 + \beta^2 + 2 \alpha \beta \cos 2 \phi}} \nonumber \\
	&= -\frac{1}{2} \langle - | \sigma_x | - \rangle = -\left\langle S_x \right\rangle_-, \label{eq:Sx} \\
	\left\langle S_y \right\rangle_+ &= \frac{1}{2} \langle + | \sigma_y | + \rangle = \frac{i}{4}(\zeta - \zeta^*) \nonumber \\
	&= \frac{1}{2 \sqrt{2}} \frac{(\alpha + \beta) \cos \phi + (\alpha - \beta) \sin \phi}{\sqrt{\alpha^2 + \beta^2 + 2 \alpha \beta \cos 2 \phi}} \nonumber \\
	&= -\frac{1}{2} \langle - | \sigma_y | - \rangle = -\left\langle S_y \right\rangle_-. \label{eq:Sy}
\end{align}
The pseudospins obtained from eqs~(\ref{eq:Sz}), (\ref{eq:Sx}), and (\ref{eq:Sy}) is marked in \cref{fig:ModelBands}(i), highlighting the rotation of the pseudospin direction.

The formation of electron and hole pockets hints that the number of charge carriers revealed in a transport measurement may be fewer than 0.5e$^-$ per interface unit cell for this system. Nevertheless, the large Rashba-Dresselhaus effect observed here in conjunction with an antiferromagnetic ground state makes this heterostructure a lucrative material for emerging antiferromagnetic spintronics \cite{BaltzRMP18}.
\subsubsection{Proximity Effect}
\begin{figure}
    \centering
    \includegraphics[scale = 0.34]{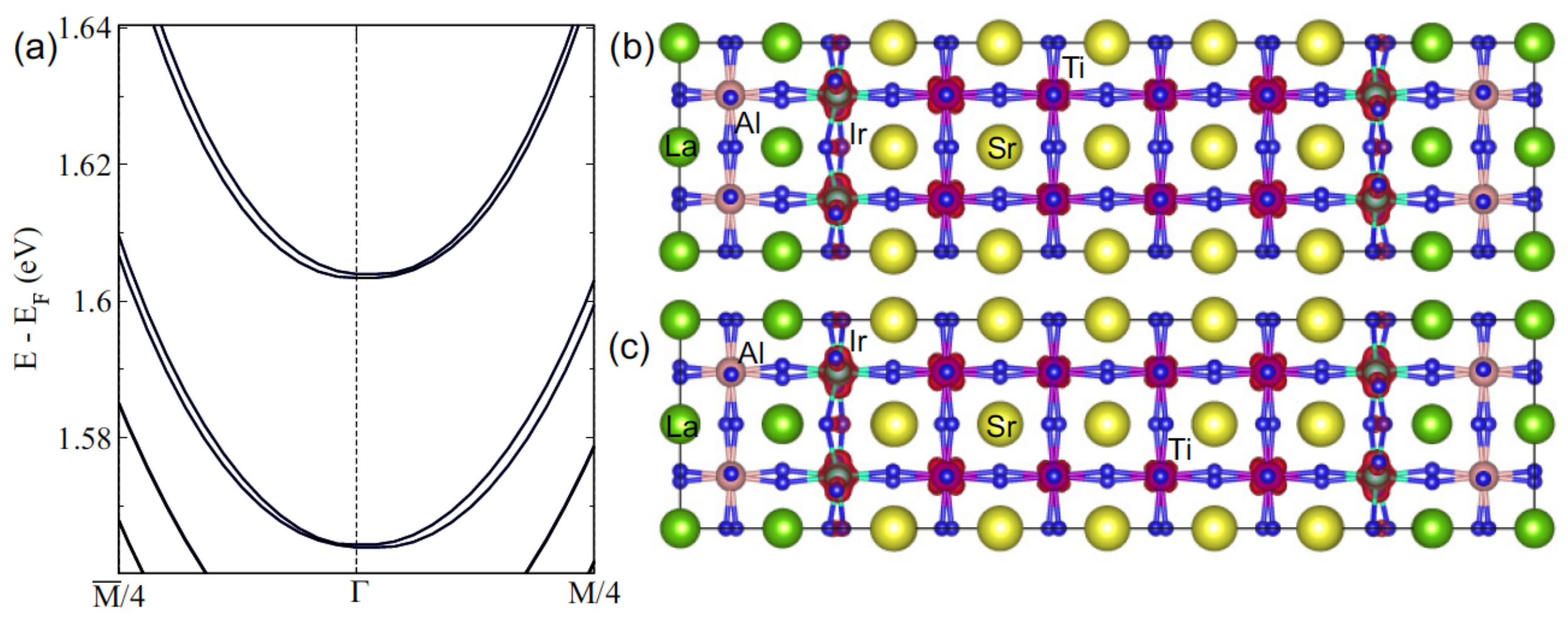}
    \caption{\label{fig:Rashba-Ti}Rashba-like effect for predominantly Ti-$3d$ bands manifested due to proximity to strong spin-orbit entangled Ir are highlighted in (a). (b) and (c) show the charge density isosurfaces corresponding to upper and lower pair of Rashba-like split bands in (a), respectively.}
\end{figure}
After confirming a large Rashba-Dresselhaus effect from Ir-$5d$ bands at LAO$|$SIO$|$STO heterojunction, we carefully examine the empty Ti-$3d$ states above the Fermi level. The relevant part of the band structure has been magnified and shown in \cref{fig:Rashba-Ti}(a), indicating much smaller, yet substantially Rashba-like split bands well above the Fermi level. Charge density isosurfaces corresponding to the upper and the lower pair of Rashba-like split bands are depicted in \cref{fig:Rashba-Ti}(b) and \cref{fig:Rashba-Ti}(c), respectively, revealing predominant Ti-$3d$ characters of the bands with some admixture of Ir-$5d$ character. These Ti- $ 3d$ bands visibly exhibit significantly more splitting than that of ref.~\onlinecite{ZhongPRB13} for LAO$|$STO. This considerably larger Rashba-like splitting from Ti-$3d$ bands can be attributed to the proximity to Ir-$5d$ states with strong spin-orbit interaction.
\section{\label{sec:discuss}Discussion}
The heterostructure of LAO$|$SIO$|$STO simulated here possesses a number of interesting properties at the ultra-thin SIO layer, namely, (i) two-dimensional conducting layer owing to charge transfer, (ii) broken inversion symmetry and strong spin-orbit interaction driven anisotropic Rashba-Dresselhaus effect, (iii) canted antiferromagnetic order, and (iv) proximity-driven Rashba-like effect. These properties make this heterostructure attractive for exploring antiferromagnetic spintronics. The strong anisotropic Rashba-Dresselhaus effect observed from our calculations in the two-dimensional conducting antiferromagnetic layer is promising for realizing spin-orbit torque. Further, the robust insulating nature of STO and LAO with significant band gaps ensures that except for the interface, the heterostructure remains insulating everywhere else, allowing for tuning the Rashba-Dresselhaus coefficients and the possible spin-orbit torque via external electric field or gate voltage over a wide range. Various spin textures may be realized in the noncollinear antiferromagnetic interface, which allows for storing and processing information, possibly at terahertz frequency and extremely low power, controlled via spin-orbit torque. Besides, the observation of enhanced Rashba-like effect in Ti-$3d$ states due to its proximity to Ir suggests opening up a new horizon in designing proximity-driven heterostructures for future technology. One of the advantages of proximity-induced spin-orbit interaction over adatom or substitution is that the host material's electronic structure is only slightly altered. Nevertheless, experimental verification of the predicted properties like the presence of a two-dimensional conducting layer, canted antiferromagnetism, and strong anisotropic Rashba-Dresselhaus effect at the interface; and hypothesized property like tunable spin-orbit torque are imperative for further progress.
\section{\label{sec:conc}Conclusion}
To conclude, we have simulated a SrIrO$_3|$SrTiO$_3$ and a LaAlO$_3|$SrIrO$_3|$SrTiO$_3$ heterostructure within the framework of density functional theory including Hubbard-$U$ correction and spin-orbit interaction. Our results suggest that the nonpolar SrIrO$_3|$SrTiO$_3$ heterostructure behaves like a canted antiferromagnetic Mott/Slater insulator with tilted IrO$_6$ octahedra. The polar$|$nonpolar heterostructure of LaAlO$_3|$SrIrO$_3|$SrTiO$_3$, where we assume one unit cell of SrIrO$_3$ sandwiched between a thin film of LaAlO$_3$ and a thick substrate of SrTiO$_3$, results in a two-dimensional conducting layer with electron and hole pockets at the interface SrIrO$_3$ unit cell. Our calculations reveal a canted antiferromagnetic ground state also for the LaAlO$_3|$SrIrO$_3|$SrTiO$_3$ heterostructure with tilted IrO$_6$ octahedra. We observe a strong anisotropic Rashba-Dresselhaus effect from the bands contributing to the conduction of charges. A physical model developed for anisotropic Rashba-Dresselhaus effect without too many parameters nicely explains our results obtained from DFT+$U$+SOC calculations, providing rich physical insight into the system and illustrating pseudospin orientation. Based on these results, we hypothesize that electric field tunable spin-orbit torque may be realized at this heterojunction, making it an excellent testbed for antiferromagnetic spintronics. Our predictions and hypotheses call for experimental verification for further progress in the field. In another crucial outcome, we find a pronounced Rashba-like effect at Ti-$3d$ empty states because of its proximity to strong spin-orbit coupled Ir atoms, provoking new designs of proximity-driven heterostructures for future technology.

\appendix
\section{\label{sec:BZ}The Brillouin zone}
\begin{figure}[b]
	\centering
	\includegraphics[scale = 1]{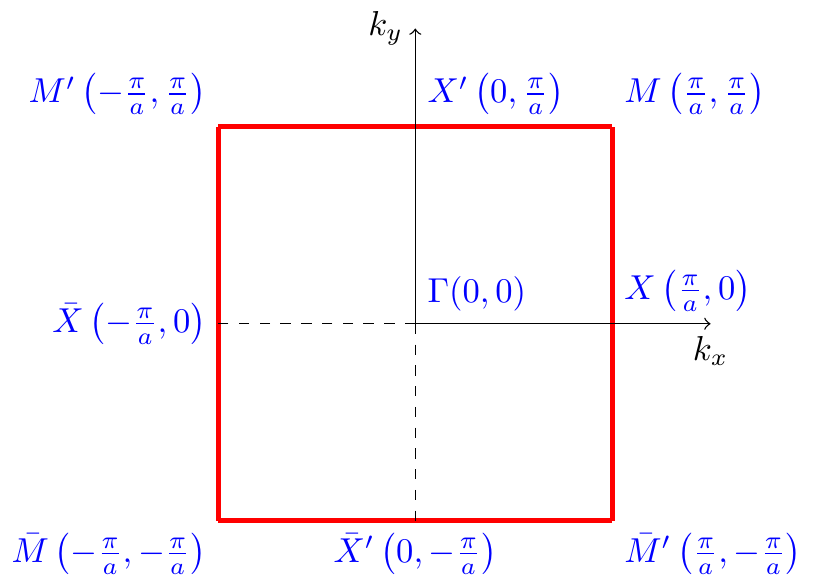}
	\caption{\label{fig:BrillouinZone}A two-dimensional Brillouin zone of square shape is depicted here. The relevant high symmetry points i.e. the $\Gamma$-point, four $X$-equivalent points, and four $M$-equivalent points along with their coordinates are marked in the figure.}
\end{figure}
The geometry of the Brillouin zone is important to understand the band dispersion of the heterostructures under consideration. Since we are interested in heterostructures of perovskite oxides where periodicity along the $c$-direction is unimportant, we can consider a two-dimensional (2D) Brillouin zone. A square lattice in the $ab$ plane would lead to a square-shaped Brillouin zone, as shown in \cref{fig:BrillouinZone}. The zone boundary is marked with thick red lines, and the momentum axes $k_x$ and $k_y$ are marked with black lines. The high symmetry points along with their coordinates are marked on the figure. It is customary to represent the coordinates in the Brillouin zone in the units of $2 \pi/a$, transforming the coordinates of $X$-point into $(0.5, 0)$, the coordinates of $M$-point into $(0.5, 0.5)$ and so on. A combination of inversion symmetry and four-fold rotational symmetry transforms the irreducible part of the Brillouin zone into a triangle with $X$, $\Gamma$, and $M$ points at the corners. However, the systems considered here do not obey any of these symmetries, making the entire Brillouin zone irreducible. The effective 2D nature of the Brillouin zone allows for the band dispersion $\varepsilon(\vec{k})$, $\vec{k} = (k_x, k_y)$ being adequately represented in a 3D figure. Further, the Fermi surface (constant energy surface) reduces to Fermi contours (constant energy contours) in this context.
\section{\label{sec:Rashba}No signature of Rashba effect in $\text{SrIrO}_3|\text{SrTiO}_3$ heterostructure}
\begin{figure}
    \centering
    \includegraphics[scale = 0.34]{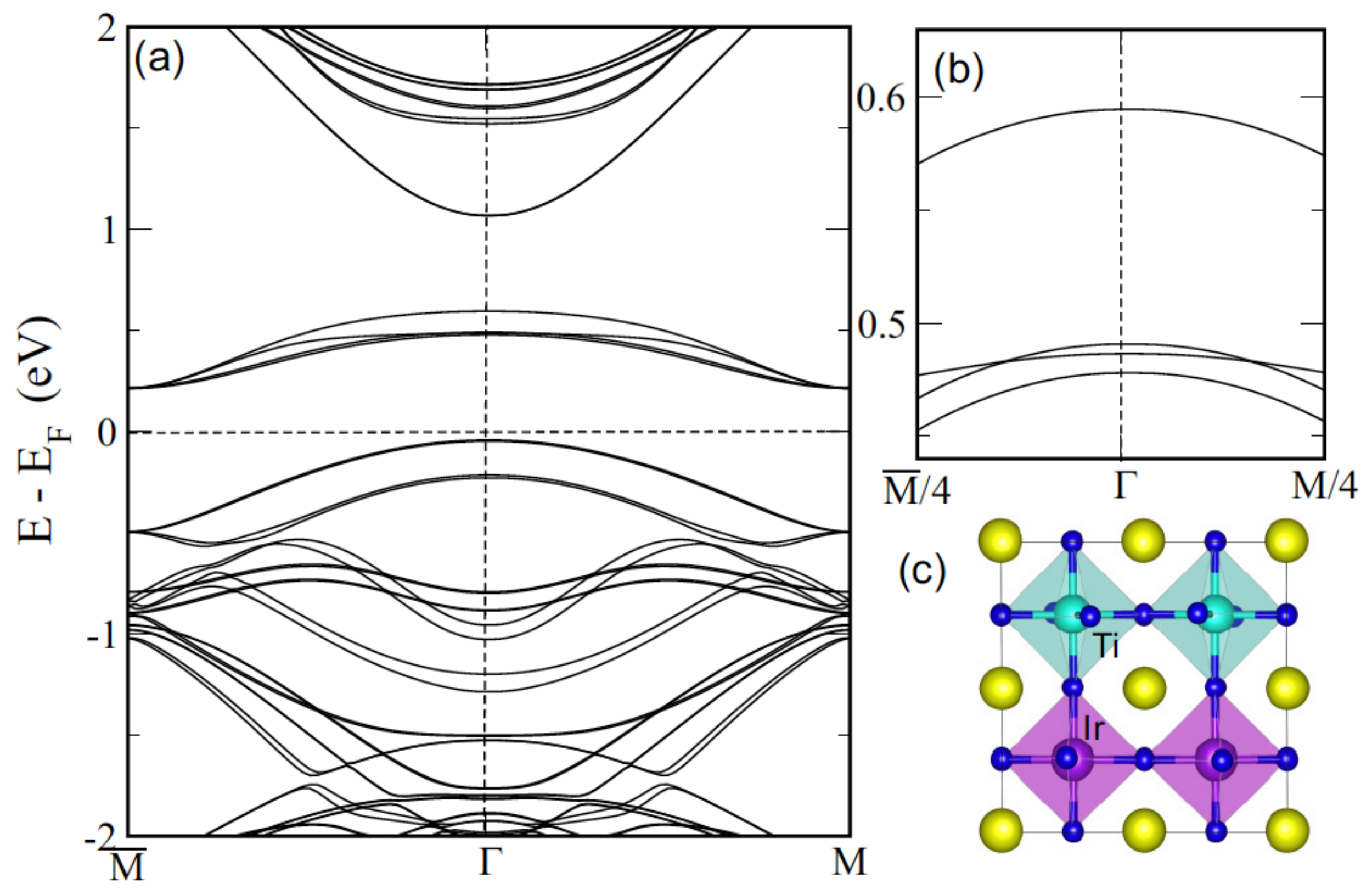}
    \caption{\label{fig:SIO-STO-Band}Subfigure (a) shows the band structure of (SrIrO$_3)_1|$(SrTiO$_3)_1$ heterostructure along $\bar{M} \to \Gamma \to M$, while (b) zooms in the Ir-$5d$ bands above the Fermi level, revealing no signature of Rashba-like effect. Subfigure (c) depicts the structure of  SrIrO$_3|$SrTiO$_3$ as viewed from $a$-axis, revealing the absence of tilt about $a$-axis.}
\end{figure}
As discussed earlier, an electric field is generated in the heterostructure due to the deposition of polar perovskite oxide thin film of LaAlO$_3$. The electrostatics of heterostructuring a polar and a nonpolar perovskite oxide has been discussed in detail in ref.~\onlinecite{GanguliPRL14}. We aim at understanding the role of this electric field for the Rashba-like effect. Therefore, we carefully examine the band structure of SrIrO$_3|$SrTiO$_3$ heterostructure calculated within the GGA+$U$+SOC method in the absence of the polar LaAlO$_3$ thin film. Our results, shown in \cref{fig:SIO-STO-Band}(a), display the bands along $\bar{M} \to \Gamma \to M$ direction, while \cref{fig:SIO-STO-Band}(b) zooms in the Ir-$5d$ bands right above the Fermi level to highlight any Rashba-like effect. We hardly observe any Rashba-like effect for this heterostructure, consistent with an experimental report of temperature-dependent weak Rashba-like effect \cite{ZhangJPSJ14}, seemingly because of the absence of any microscopic electric field. However, an externally applied electric field along the $c$-direction may introduce substantial Rashba-like effect in this system, owing to the spin-orbit entangled Ir-$5d$ states. Such an electric field may also transform the heterostructure into a conducting one.
\begin{acknowledgements}
Financial supports from SERB India through grant number ECR/2016/001004 and the use of high-performance computing facility at IISER Bhopal are gratefully acknowledged.
\end{acknowledgements}
%

\begin{thebibliography}{37}%
\makeatletter
\providecommand \@ifxundefined [1]{%
 \@ifx{#1\undefined}
}%
\providecommand \@ifnum [1]{%
 \ifnum #1\expandafter \@firstoftwo
 \else \expandafter \@secondoftwo
 \fi
}%
\providecommand \@ifx [1]{%
 \ifx #1\expandafter \@firstoftwo
 \else \expandafter \@secondoftwo
 \fi
}%
\providecommand \natexlab [1]{#1}%
\providecommand \enquote  [1]{``#1''}%
\providecommand \bibnamefont  [1]{#1}%
\providecommand \bibfnamefont [1]{#1}%
\providecommand \citenamefont [1]{#1}%
\providecommand \href@noop [0]{\@secondoftwo}%
\providecommand \href [0]{\begingroup \@sanitize@url \@href}%
\providecommand \@href[1]{\@@startlink{#1}\@@href}%
\providecommand \@@href[1]{\endgroup#1\@@endlink}%
\providecommand \@sanitize@url [0]{\catcode `\\12\catcode `\$12\catcode
  `\&12\catcode `\#12\catcode `\^12\catcode `\_12\catcode `\%12\relax}%
\providecommand \@@startlink[1]{}%
\providecommand \@@endlink[0]{}%
\providecommand \url  [0]{\begingroup\@sanitize@url \@url }%
\providecommand \@url [1]{\endgroup\@href {#1}{\urlprefix }}%
\providecommand \urlprefix  [0]{URL }%
\providecommand \Eprint [0]{\href }%
\providecommand \doibase [0]{https://doi.org/}%
\providecommand \selectlanguage [0]{\@gobble}%
\providecommand \bibinfo  [0]{\@secondoftwo}%
\providecommand \bibfield  [0]{\@secondoftwo}%
\providecommand \translation [1]{[#1]}%
\providecommand \BibitemOpen [0]{}%
\providecommand \bibitemStop [0]{}%
\providecommand \bibitemNoStop [0]{.\EOS\space}%
\providecommand \EOS [0]{\spacefactor3000\relax}%
\providecommand \BibitemShut  [1]{\csname bibitem#1\endcsname}%
\let\auto@bib@innerbib\@empty
\bibitem [{\citenamefont {Baltz}\ \emph {et~al.}(2018)\citenamefont {Baltz},
  \citenamefont {Manchon}, \citenamefont {Tsoi}, \citenamefont {Moriyama},
  \citenamefont {Ono},\ and\ \citenamefont {Tserkovnyak}}]{BaltzRMP18}%
  \BibitemOpen
  \bibfield  {author} {\bibinfo {author} {\bibfnamefont {V.}~\bibnamefont
  {Baltz}}, \bibinfo {author} {\bibfnamefont {A.}~\bibnamefont {Manchon}},
  \bibinfo {author} {\bibfnamefont {M.}~\bibnamefont {Tsoi}}, \bibinfo {author}
  {\bibfnamefont {T.}~\bibnamefont {Moriyama}}, \bibinfo {author}
  {\bibfnamefont {T.}~\bibnamefont {Ono}},\ and\ \bibinfo {author}
  {\bibfnamefont {Y.}~\bibnamefont {Tserkovnyak}},\ }\bibfield  {title}
  {\bibinfo {title} {{Antiferromagnetic spintronics}},\ }\href
  {https://doi.org/10.1103/RevModPhys.90.015005} {\bibfield  {journal}
  {\bibinfo  {journal} {Rev. Mod. Phys.}\ }\textbf {\bibinfo {volume} {90}},\
  \bibinfo {pages} {015005} (\bibinfo {year} {2018})}\BibitemShut {NoStop}%
\bibitem [{\citenamefont {{\v{Z}}elezn{\'{y}}}\ \emph
  {et~al.}(2018)\citenamefont {{\v{Z}}elezn{\'{y}}}, \citenamefont {Wadley},
  \citenamefont {Olejn{\'{i}}k}, \citenamefont {Hoffmann},\ and\ \citenamefont
  {Ohno}}]{ZeleznyNP18}%
  \BibitemOpen
  \bibfield  {author} {\bibinfo {author} {\bibfnamefont {J.}~\bibnamefont
  {{\v{Z}}elezn{\'{y}}}}, \bibinfo {author} {\bibfnamefont {P.}~\bibnamefont
  {Wadley}}, \bibinfo {author} {\bibfnamefont {K.}~\bibnamefont
  {Olejn{\'{i}}k}}, \bibinfo {author} {\bibfnamefont {A.}~\bibnamefont
  {Hoffmann}},\ and\ \bibinfo {author} {\bibfnamefont {H.}~\bibnamefont
  {Ohno}},\ }\bibfield  {title} {\bibinfo {title} {{Spin transport and spin
  torque in antiferromagnetic devices}},\ }\href
  {https://doi.org/10.1038/s41567-018-0062-7} {\bibfield  {journal} {\bibinfo
  {journal} {Nat. Phys.}\ }\textbf {\bibinfo {volume} {14}},\ \bibinfo {pages}
  {220} (\bibinfo {year} {2018})}\BibitemShut {NoStop}%
\bibitem [{\citenamefont {Gomonay}\ \emph {et~al.}(2018)\citenamefont
  {Gomonay}, \citenamefont {Baltz}, \citenamefont {Brataas},\ and\
  \citenamefont {Tserkovnyak}}]{GomonayNP18}%
  \BibitemOpen
  \bibfield  {author} {\bibinfo {author} {\bibfnamefont {O.}~\bibnamefont
  {Gomonay}}, \bibinfo {author} {\bibfnamefont {V.}~\bibnamefont {Baltz}},
  \bibinfo {author} {\bibfnamefont {A.}~\bibnamefont {Brataas}},\ and\ \bibinfo
  {author} {\bibfnamefont {Y.}~\bibnamefont {Tserkovnyak}},\ }\bibfield
  {title} {\bibinfo {title} {{Antiferromagnetic spin textures and dynamics}},\
  }\href {https://doi.org/10.1038/s41567-018-0049-4} {\bibfield  {journal}
  {\bibinfo  {journal} {Nat. Phys.}\ }\textbf {\bibinfo {volume} {14}},\
  \bibinfo {pages} {213} (\bibinfo {year} {2018})}\BibitemShut {NoStop}%
\bibitem [{\citenamefont {Jungwirth}\ \emph {et~al.}(2016)\citenamefont
  {Jungwirth}, \citenamefont {Marti}, \citenamefont {Wadley},\ and\
  \citenamefont {Wunderlich}}]{JungwirthNN16}%
  \BibitemOpen
  \bibfield  {author} {\bibinfo {author} {\bibfnamefont {T.}~\bibnamefont
  {Jungwirth}}, \bibinfo {author} {\bibfnamefont {X.}~\bibnamefont {Marti}},
  \bibinfo {author} {\bibfnamefont {P.}~\bibnamefont {Wadley}},\ and\ \bibinfo
  {author} {\bibfnamefont {J.}~\bibnamefont {Wunderlich}},\ }\bibfield  {title}
  {\bibinfo {title} {{Antiferromagnetic spintronics}},\ }\href
  {https://doi.org/10.1038/nnano.2016.18} {\bibfield  {journal} {\bibinfo
  {journal} {Nat. Nanotech.}\ }\textbf {\bibinfo {volume} {11}},\ \bibinfo
  {pages} {231} (\bibinfo {year} {2016})}\BibitemShut {NoStop}%
\bibitem [{\citenamefont {Olejn{\'{i}}k}\ \emph {et~al.}(2018)\citenamefont
  {Olejn{\'{i}}k}, \citenamefont {Seifert}, \citenamefont {Ka{\v{s}}par},
  \citenamefont {Nov{\'{a}}k}, \citenamefont {Wadley}, \citenamefont {Campion},
  \citenamefont {Baumgartner}, \citenamefont {Gambardella}, \citenamefont
  {N{\v{e}}mec}, \citenamefont {Wunderlich}, \citenamefont {Sinova},
  \citenamefont {Ku{\v{z}}el}, \citenamefont {M{\"{u}}ller}, \citenamefont
  {Kampfrath},\ and\ \citenamefont {Jungwirth}}]{OlejnikSA18}%
  \BibitemOpen
  \bibfield  {author} {\bibinfo {author} {\bibfnamefont {K.}~\bibnamefont
  {Olejn{\'{i}}k}}, \bibinfo {author} {\bibfnamefont {T.}~\bibnamefont
  {Seifert}}, \bibinfo {author} {\bibfnamefont {Z.}~\bibnamefont
  {Ka{\v{s}}par}}, \bibinfo {author} {\bibfnamefont {V.}~\bibnamefont
  {Nov{\'{a}}k}}, \bibinfo {author} {\bibfnamefont {P.}~\bibnamefont {Wadley}},
  \bibinfo {author} {\bibfnamefont {R.~P.}\ \bibnamefont {Campion}}, \bibinfo
  {author} {\bibfnamefont {M.}~\bibnamefont {Baumgartner}}, \bibinfo {author}
  {\bibfnamefont {P.}~\bibnamefont {Gambardella}}, \bibinfo {author}
  {\bibfnamefont {P.}~\bibnamefont {N{\v{e}}mec}}, \bibinfo {author}
  {\bibfnamefont {J.}~\bibnamefont {Wunderlich}}, \bibinfo {author}
  {\bibfnamefont {J.}~\bibnamefont {Sinova}}, \bibinfo {author} {\bibfnamefont
  {P.}~\bibnamefont {Ku{\v{z}}el}}, \bibinfo {author} {\bibfnamefont
  {M.}~\bibnamefont {M{\"{u}}ller}}, \bibinfo {author} {\bibfnamefont
  {T.}~\bibnamefont {Kampfrath}},\ and\ \bibinfo {author} {\bibfnamefont
  {T.}~\bibnamefont {Jungwirth}},\ }\bibfield  {title} {\bibinfo {title}
  {{Terahertz electrical writing speed in an antiferromagnetic memory}},\
  }\href {https://doi.org/10.1126/sciadv.aar3566} {\bibfield  {journal}
  {\bibinfo  {journal} {Sci. Adv.}\ }\textbf {\bibinfo {volume} {4}},\ \bibinfo
  {pages} {eaar3566} (\bibinfo {year} {2018})}\BibitemShut {NoStop}%
\bibitem [{\citenamefont {Kim}\ \emph {et~al.}(2008)\citenamefont {Kim},
  \citenamefont {Jin}, \citenamefont {Moon}, \citenamefont {Kim}, \citenamefont
  {Park}, \citenamefont {Leem}, \citenamefont {Yu}, \citenamefont {Noh},
  \citenamefont {Kim}, \citenamefont {Oh}, \citenamefont {Park}, \citenamefont
  {Durairaj}, \citenamefont {Cao},\ and\ \citenamefont {Rotenberg}}]{KimPRL08}%
  \BibitemOpen
  \bibfield  {author} {\bibinfo {author} {\bibfnamefont {B.~J.}\ \bibnamefont
  {Kim}}, \bibinfo {author} {\bibfnamefont {H.}~\bibnamefont {Jin}}, \bibinfo
  {author} {\bibfnamefont {S.~J.}\ \bibnamefont {Moon}}, \bibinfo {author}
  {\bibfnamefont {J.-Y.}\ \bibnamefont {Kim}}, \bibinfo {author} {\bibfnamefont
  {B.-G.}\ \bibnamefont {Park}}, \bibinfo {author} {\bibfnamefont {C.~S.}\
  \bibnamefont {Leem}}, \bibinfo {author} {\bibfnamefont {J.}~\bibnamefont
  {Yu}}, \bibinfo {author} {\bibfnamefont {T.~W.}\ \bibnamefont {Noh}},
  \bibinfo {author} {\bibfnamefont {C.}~\bibnamefont {Kim}}, \bibinfo {author}
  {\bibfnamefont {S.-J.}\ \bibnamefont {Oh}}, \bibinfo {author} {\bibfnamefont
  {J.-H.}\ \bibnamefont {Park}}, \bibinfo {author} {\bibfnamefont
  {V.}~\bibnamefont {Durairaj}}, \bibinfo {author} {\bibfnamefont
  {G.}~\bibnamefont {Cao}},\ and\ \bibinfo {author} {\bibfnamefont
  {E.}~\bibnamefont {Rotenberg}},\ }\bibfield  {title} {\bibinfo {title}
  {{Novel $J_\text{eff} = 1/2$ Mott State Induced by Relativistic Spin-Orbit
  Coupling in Sr$_2$IrO$_4$}},\ }\href
  {https://doi.org/10.1103/PhysRevLett.101.076402} {\bibfield  {journal}
  {\bibinfo  {journal} {Phys. Rev. Lett.}\ }\textbf {\bibinfo {volume} {101}},\
  \bibinfo {pages} {076402} (\bibinfo {year} {2008})}\BibitemShut {NoStop}%
\bibitem [{\citenamefont {Modic}\ \emph {et~al.}(2014)\citenamefont {Modic},
  \citenamefont {Smidt}, \citenamefont {Kimchi}, \citenamefont {Breznay},
  \citenamefont {Biffin}, \citenamefont {Choi}, \citenamefont {Johnson},
  \citenamefont {Coldea}, \citenamefont {Watkins-Curry}, \citenamefont
  {McCandless}, \citenamefont {Chan}, \citenamefont {Gandara}, \citenamefont
  {Islam}, \citenamefont {Vishwanath}, \citenamefont {Shekhter}, \citenamefont
  {McDonald},\ and\ \citenamefont {Analytis}}]{ModicNC14}%
  \BibitemOpen
  \bibfield  {author} {\bibinfo {author} {\bibfnamefont {K.~A.}\ \bibnamefont
  {Modic}}, \bibinfo {author} {\bibfnamefont {T.~E.}\ \bibnamefont {Smidt}},
  \bibinfo {author} {\bibfnamefont {I.}~\bibnamefont {Kimchi}}, \bibinfo
  {author} {\bibfnamefont {N.~P.}\ \bibnamefont {Breznay}}, \bibinfo {author}
  {\bibfnamefont {A.}~\bibnamefont {Biffin}}, \bibinfo {author} {\bibfnamefont
  {S.}~\bibnamefont {Choi}}, \bibinfo {author} {\bibfnamefont {R.~D.}\
  \bibnamefont {Johnson}}, \bibinfo {author} {\bibfnamefont {R.}~\bibnamefont
  {Coldea}}, \bibinfo {author} {\bibfnamefont {P.}~\bibnamefont
  {Watkins-Curry}}, \bibinfo {author} {\bibfnamefont {G.~T.}\ \bibnamefont
  {McCandless}}, \bibinfo {author} {\bibfnamefont {J.~Y.}\ \bibnamefont
  {Chan}}, \bibinfo {author} {\bibfnamefont {F.}~\bibnamefont {Gandara}},
  \bibinfo {author} {\bibfnamefont {Z.}~\bibnamefont {Islam}}, \bibinfo
  {author} {\bibfnamefont {A.}~\bibnamefont {Vishwanath}}, \bibinfo {author}
  {\bibfnamefont {A.}~\bibnamefont {Shekhter}}, \bibinfo {author}
  {\bibfnamefont {R.~D.}\ \bibnamefont {McDonald}},\ and\ \bibinfo {author}
  {\bibfnamefont {J.~G.}\ \bibnamefont {Analytis}},\ }\bibfield  {title}
  {\bibinfo {title} {{Realization of a three-dimensional spin–anisotropic
  harmonic honeycomb iridate}},\ }\href {https://doi.org/10.1038/ncomms5203}
  {\bibfield  {journal} {\bibinfo  {journal} {Nat. Commun.}\ }\textbf {\bibinfo
  {volume} {5}},\ \bibinfo {pages} {4203} (\bibinfo {year} {2014})}\BibitemShut
  {NoStop}%
\bibitem [{\citenamefont {Chakraborty}(2018)}]{ChakrabortyPRB18}%
  \BibitemOpen
  \bibfield  {author} {\bibinfo {author} {\bibfnamefont {J.}~\bibnamefont
  {Chakraborty}},\ }\bibfield  {title} {\bibinfo {title} {{Interplay of
  covalency, spin-orbit coupling, and geometric frustration in the d$^{3.5}$
  system Ba$_3$LiIr$_2$O$_9$}},\ }\href
  {https://doi.org/10.1103/PhysRevB.97.235147} {\bibfield  {journal} {\bibinfo
  {journal} {Phys. Rev. B}\ }\textbf {\bibinfo {volume} {97}},\ \bibinfo
  {pages} {235147} (\bibinfo {year} {2018})}\BibitemShut {NoStop}%
\bibitem [{\citenamefont {Matsuno}\ \emph {et~al.}(2015)\citenamefont
  {Matsuno}, \citenamefont {Ihara}, \citenamefont {Yamamura}, \citenamefont
  {Wadati}, \citenamefont {Ishii}, \citenamefont {Shankar}, \citenamefont
  {Kee},\ and\ \citenamefont {Takagi}}]{MatsunoPRL15}%
  \BibitemOpen
  \bibfield  {author} {\bibinfo {author} {\bibfnamefont {J.}~\bibnamefont
  {Matsuno}}, \bibinfo {author} {\bibfnamefont {K.}~\bibnamefont {Ihara}},
  \bibinfo {author} {\bibfnamefont {S.}~\bibnamefont {Yamamura}}, \bibinfo
  {author} {\bibfnamefont {H.}~\bibnamefont {Wadati}}, \bibinfo {author}
  {\bibfnamefont {K.}~\bibnamefont {Ishii}}, \bibinfo {author} {\bibfnamefont
  {V.~V.}\ \bibnamefont {Shankar}}, \bibinfo {author} {\bibfnamefont {H.-Y.}\
  \bibnamefont {Kee}},\ and\ \bibinfo {author} {\bibfnamefont {H.}~\bibnamefont
  {Takagi}},\ }\bibfield  {title} {\bibinfo {title} {{Engineering a
  Spin-Orbital Magnetic Insulator by Tailoring Superlattices}},\ }\href
  {https://doi.org/10.1103/PhysRevLett.114.247209} {\bibfield  {journal}
  {\bibinfo  {journal} {Phys. Rev. Lett.}\ }\textbf {\bibinfo {volume} {114}},\
  \bibinfo {pages} {247209} (\bibinfo {year} {2015})}\BibitemShut {NoStop}%
\bibitem [{\citenamefont {Groenendijk}\ \emph {et~al.}(2017)\citenamefont
  {Groenendijk}, \citenamefont {Autieri}, \citenamefont {Girovsky},
  \citenamefont {Martinez-Velarte}, \citenamefont {Manca}, \citenamefont
  {Mattoni}, \citenamefont {Monteiro}, \citenamefont {Gauquelin}, \citenamefont
  {Verbeeck}, \citenamefont {Otte}, \citenamefont {Gabay}, \citenamefont
  {Picozzi},\ and\ \citenamefont {Caviglia}}]{GroenendijkPRL17}%
  \BibitemOpen
  \bibfield  {author} {\bibinfo {author} {\bibfnamefont {D.~J.}\ \bibnamefont
  {Groenendijk}}, \bibinfo {author} {\bibfnamefont {C.}~\bibnamefont
  {Autieri}}, \bibinfo {author} {\bibfnamefont {J.}~\bibnamefont {Girovsky}},
  \bibinfo {author} {\bibfnamefont {M.~C.}\ \bibnamefont {Martinez-Velarte}},
  \bibinfo {author} {\bibfnamefont {N.}~\bibnamefont {Manca}}, \bibinfo
  {author} {\bibfnamefont {G.}~\bibnamefont {Mattoni}}, \bibinfo {author}
  {\bibfnamefont {A.~M. R. V.~L.}\ \bibnamefont {Monteiro}}, \bibinfo {author}
  {\bibfnamefont {N.}~\bibnamefont {Gauquelin}}, \bibinfo {author}
  {\bibfnamefont {J.}~\bibnamefont {Verbeeck}}, \bibinfo {author}
  {\bibfnamefont {A.~F.}\ \bibnamefont {Otte}}, \bibinfo {author}
  {\bibfnamefont {M.}~\bibnamefont {Gabay}}, \bibinfo {author} {\bibfnamefont
  {S.}~\bibnamefont {Picozzi}},\ and\ \bibinfo {author} {\bibfnamefont {A.~D.}\
  \bibnamefont {Caviglia}},\ }\bibfield  {title} {\bibinfo {title} {{Spin-Orbit
  Semimetal SrIrO$_3$ in the Two-Dimensional Limit}},\ }\href
  {https://doi.org/10.1103/PhysRevLett.119.256403} {\bibfield  {journal}
  {\bibinfo  {journal} {Phys. Rev. Lett.}\ }\textbf {\bibinfo {volume} {119}},\
  \bibinfo {pages} {256403} (\bibinfo {year} {2017})}\BibitemShut {NoStop}%
\bibitem [{\citenamefont {Ohuchi}\ \emph {et~al.}(2018)\citenamefont {Ohuchi},
  \citenamefont {Matsuno}, \citenamefont {Ogawa}, \citenamefont {Kozuka},
  \citenamefont {Uchida}, \citenamefont {Tokura},\ and\ \citenamefont
  {Kawasaki}}]{OhuchiNC18}%
  \BibitemOpen
  \bibfield  {author} {\bibinfo {author} {\bibfnamefont {Y.}~\bibnamefont
  {Ohuchi}}, \bibinfo {author} {\bibfnamefont {J.}~\bibnamefont {Matsuno}},
  \bibinfo {author} {\bibfnamefont {N.}~\bibnamefont {Ogawa}}, \bibinfo
  {author} {\bibfnamefont {Y.}~\bibnamefont {Kozuka}}, \bibinfo {author}
  {\bibfnamefont {M.}~\bibnamefont {Uchida}}, \bibinfo {author} {\bibfnamefont
  {Y.}~\bibnamefont {Tokura}},\ and\ \bibinfo {author} {\bibfnamefont
  {M.}~\bibnamefont {Kawasaki}},\ }\bibfield  {title} {\bibinfo {title}
  {{Electric-field control of anomalous and topological Hall effects in oxide
  bilayer thin films}},\ }\href {https://doi.org/10.1038/s41467-017-02629-3}
  {\bibfield  {journal} {\bibinfo  {journal} {Nat. Commun.}\ }\textbf {\bibinfo
  {volume} {9}},\ \bibinfo {pages} {213} (\bibinfo {year} {2018})}\BibitemShut
  {NoStop}%
\bibitem [{\citenamefont {Bhandari}\ and\ \citenamefont
  {Satpathy}(2018)}]{BhandariPRB18}%
  \BibitemOpen
  \bibfield  {author} {\bibinfo {author} {\bibfnamefont {C.}~\bibnamefont
  {Bhandari}}\ and\ \bibinfo {author} {\bibfnamefont {S.}~\bibnamefont
  {Satpathy}},\ }\bibfield  {title} {\bibinfo {title} {{Spin-orbital entangled
  two-dimensional electron gas at the LaAlO$_3$/Sr$_2$IrO$_4$ interface}},\
  }\href {https://doi.org/10.1103/PhysRevB.98.041303} {\bibfield  {journal}
  {\bibinfo  {journal} {Phys. Rev. B}\ }\textbf {\bibinfo {volume} {98}},\
  \bibinfo {pages} {041303} (\bibinfo {year} {2018})}\BibitemShut {NoStop}%
\bibitem [{\citenamefont {Everhardt}\ \emph {et~al.}(2019)\citenamefont
  {Everhardt}, \citenamefont {DC}, \citenamefont {Huang}, \citenamefont
  {Sayed}, \citenamefont {Gosavi}, \citenamefont {Tang}, \citenamefont {Lin},
  \citenamefont {Manipatruni}, \citenamefont {Young}, \citenamefont {Datta},
  \citenamefont {Wang},\ and\ \citenamefont {Ramesh}}]{EverhardtPRM19}%
  \BibitemOpen
  \bibfield  {author} {\bibinfo {author} {\bibfnamefont {A.~S.}\ \bibnamefont
  {Everhardt}}, \bibinfo {author} {\bibfnamefont {M.}~\bibnamefont {DC}},
  \bibinfo {author} {\bibfnamefont {X.}~\bibnamefont {Huang}}, \bibinfo
  {author} {\bibfnamefont {S.}~\bibnamefont {Sayed}}, \bibinfo {author}
  {\bibfnamefont {T.~A.}\ \bibnamefont {Gosavi}}, \bibinfo {author}
  {\bibfnamefont {Y.}~\bibnamefont {Tang}}, \bibinfo {author} {\bibfnamefont
  {C.-C.}\ \bibnamefont {Lin}}, \bibinfo {author} {\bibfnamefont
  {S.}~\bibnamefont {Manipatruni}}, \bibinfo {author} {\bibfnamefont {I.~A.}\
  \bibnamefont {Young}}, \bibinfo {author} {\bibfnamefont {S.}~\bibnamefont
  {Datta}}, \bibinfo {author} {\bibfnamefont {J.-P.}\ \bibnamefont {Wang}},\
  and\ \bibinfo {author} {\bibfnamefont {R.}~\bibnamefont {Ramesh}},\
  }\bibfield  {title} {\bibinfo {title} {{Tunable charge to spin conversion in
  strontium iridate thin films}},\ }\href
  {https://doi.org/10.1103/PhysRevMaterials.3.051201} {\bibfield  {journal}
  {\bibinfo  {journal} {Phys. Rev. Materials}\ }\textbf {\bibinfo {volume}
  {3}},\ \bibinfo {pages} {051201} (\bibinfo {year} {2019})}\BibitemShut
  {NoStop}%
\bibitem [{\citenamefont {Wang}\ \emph {et~al.}(2019)\citenamefont {Wang},
  \citenamefont {Meng}, \citenamefont {Zhang}, \citenamefont {Hou},
  \citenamefont {Finley}, \citenamefont {Han}, \citenamefont {Yang},\ and\
  \citenamefont {Liu}}]{WangAPL19}%
  \BibitemOpen
  \bibfield  {author} {\bibinfo {author} {\bibfnamefont {H.}~\bibnamefont
  {Wang}}, \bibinfo {author} {\bibfnamefont {K.-Y.}\ \bibnamefont {Meng}},
  \bibinfo {author} {\bibfnamefont {P.}~\bibnamefont {Zhang}}, \bibinfo
  {author} {\bibfnamefont {J.~T.}\ \bibnamefont {Hou}}, \bibinfo {author}
  {\bibfnamefont {J.}~\bibnamefont {Finley}}, \bibinfo {author} {\bibfnamefont
  {J.}~\bibnamefont {Han}}, \bibinfo {author} {\bibfnamefont {F.}~\bibnamefont
  {Yang}},\ and\ \bibinfo {author} {\bibfnamefont {L.}~\bibnamefont {Liu}},\
  }\bibfield  {title} {\bibinfo {title} {{Large spin-orbit torque observed in
  epitaxial SrIrO$_3$ thin films}},\ }\href {https://doi.org/10.1063/1.5097699}
  {\bibfield  {journal} {\bibinfo  {journal} {Appl. Phys. Lett.}\ }\textbf
  {\bibinfo {volume} {114}},\ \bibinfo {pages} {232406} (\bibinfo {year}
  {2019})}\BibitemShut {NoStop}%
\bibitem [{\citenamefont {Ganguli}\ and\ \citenamefont
  {Kelly}(2014)}]{GanguliPRL14}%
  \BibitemOpen
  \bibfield  {author} {\bibinfo {author} {\bibfnamefont {N.}~\bibnamefont
  {Ganguli}}\ and\ \bibinfo {author} {\bibfnamefont {P.~J.}\ \bibnamefont
  {Kelly}},\ }\bibfield  {title} {\bibinfo {title} {{Tuning Ferromagnetism at
  Interfaces between Insulating Perovskite Oxides}},\ }\href
  {https://doi.org/10.1103/PhysRevLett.113.127201} {\bibfield  {journal}
  {\bibinfo  {journal} {Phys. Rev. Lett.}\ }\textbf {\bibinfo {volume} {113}},\
  \bibinfo {pages} {127201} (\bibinfo {year} {2014})}\BibitemShut {NoStop}%
\bibitem [{\citenamefont {Bl{\"{o}}chl}(1994)}]{paw}%
  \BibitemOpen
  \bibfield  {author} {\bibinfo {author} {\bibfnamefont {P.~E.}\ \bibnamefont
  {Bl{\"{o}}chl}},\ }\bibfield  {title} {\bibinfo {title} {{Projector
  augmented-wave method}},\ }\href {https://doi.org/10.1103/PhysRevB.50.17953}
  {\bibfield  {journal} {\bibinfo  {journal} {Phys. Rev. B}\ }\textbf {\bibinfo
  {volume} {50}},\ \bibinfo {pages} {17953} (\bibinfo {year}
  {1994})}\BibitemShut {NoStop}%
\bibitem [{\citenamefont {Kresse}\ and\ \citenamefont
  {Hafner}(1993)}]{KressePRB93}%
  \BibitemOpen
  \bibfield  {author} {\bibinfo {author} {\bibfnamefont {G.}~\bibnamefont
  {Kresse}}\ and\ \bibinfo {author} {\bibfnamefont {J.}~\bibnamefont
  {Hafner}},\ }\bibfield  {title} {\bibinfo {title} {{Ab initio molecular
  dynamics for liquid metals}},\ }\href
  {https://doi.org/10.1103/PhysRevB.47.558} {\bibfield  {journal} {\bibinfo
  {journal} {Phys. Rev. B}\ }\textbf {\bibinfo {volume} {47}},\ \bibinfo
  {pages} {558} (\bibinfo {year} {1993})}\BibitemShut {NoStop}%
\bibitem [{\citenamefont {Kresse}\ and\ \citenamefont {Joubert}(1999)}]{vasp2}%
  \BibitemOpen
  \bibfield  {author} {\bibinfo {author} {\bibfnamefont {G.}~\bibnamefont
  {Kresse}}\ and\ \bibinfo {author} {\bibfnamefont {D.}~\bibnamefont
  {Joubert}},\ }\bibfield  {title} {\bibinfo {title} {{From ultrasoft
  pseudopotentials to the projector augmented-wave method}},\ }\href
  {https://doi.org/10.1103/PhysRevB.59.1758} {\bibfield  {journal} {\bibinfo
  {journal} {Phys. Rev. B}\ }\textbf {\bibinfo {volume} {59}},\ \bibinfo
  {pages} {1758} (\bibinfo {year} {1999})}\BibitemShut {NoStop}%
\bibitem [{\citenamefont {Perdew}\ \emph {et~al.}(1996)\citenamefont {Perdew},
  \citenamefont {Burke},\ and\ \citenamefont {Ernzerhof}}]{pbe}%
  \BibitemOpen
  \bibfield  {author} {\bibinfo {author} {\bibfnamefont {J.~P.}\ \bibnamefont
  {Perdew}}, \bibinfo {author} {\bibfnamefont {K.}~\bibnamefont {Burke}},\ and\
  \bibinfo {author} {\bibfnamefont {M.}~\bibnamefont {Ernzerhof}},\ }\bibfield
  {title} {\bibinfo {title} {{Generalized Gradient Approximation Made
  Simple}},\ }\href {https://doi.org/10.1103/PhysRevLett.77.3865} {\bibfield
  {journal} {\bibinfo  {journal} {Phys. Rev. Lett.}\ }\textbf {\bibinfo
  {volume} {77}},\ \bibinfo {pages} {3865} (\bibinfo {year}
  {1996})}\BibitemShut {NoStop}%
\bibitem [{\citenamefont {Ceperley}\ and\ \citenamefont {Alder}(1980)}]{ldaCA}%
  \BibitemOpen
  \bibfield  {author} {\bibinfo {author} {\bibfnamefont {D.~M.}\ \bibnamefont
  {Ceperley}}\ and\ \bibinfo {author} {\bibfnamefont {B.~J.}\ \bibnamefont
  {Alder}},\ }\bibfield  {title} {\bibinfo {title} {{Ground State of the
  Electron Gas by a Stochastic Method}},\ }\href
  {https://doi.org/10.1103/PhysRevLett.45.566} {\bibfield  {journal} {\bibinfo
  {journal} {Phys. Rev. Lett.}\ }\textbf {\bibinfo {volume} {45}},\ \bibinfo
  {pages} {566} (\bibinfo {year} {1980})}\BibitemShut {NoStop}%
\bibitem [{\citenamefont {Perdew}\ and\ \citenamefont
  {Zunger}(1981)}]{PerdewPRB81}%
  \BibitemOpen
  \bibfield  {author} {\bibinfo {author} {\bibfnamefont {J.~P.}\ \bibnamefont
  {Perdew}}\ and\ \bibinfo {author} {\bibfnamefont {A.}~\bibnamefont
  {Zunger}},\ }\bibfield  {title} {\bibinfo {title} {{Self-interaction
  correction to density-functional approximations for many-electron systems}},\
  }\href {https://doi.org/10.1103/PhysRevB.23.5048} {\bibfield  {journal}
  {\bibinfo  {journal} {Phys. Rev. B}\ }\textbf {\bibinfo {volume} {23}},\
  \bibinfo {pages} {5048} (\bibinfo {year} {1981})}\BibitemShut {NoStop}%
\bibitem [{\citenamefont {Dudarev}\ \emph {et~al.}(1998)\citenamefont
  {Dudarev}, \citenamefont {Botton}, \citenamefont {Savrasov}, \citenamefont
  {Humphreys},\ and\ \citenamefont {Sutton}}]{DudarevPRB98}%
  \BibitemOpen
  \bibfield  {author} {\bibinfo {author} {\bibfnamefont {S.~L.}\ \bibnamefont
  {Dudarev}}, \bibinfo {author} {\bibfnamefont {G.~A.}\ \bibnamefont {Botton}},
  \bibinfo {author} {\bibfnamefont {S.~Y.}\ \bibnamefont {Savrasov}}, \bibinfo
  {author} {\bibfnamefont {C.~J.}\ \bibnamefont {Humphreys}},\ and\ \bibinfo
  {author} {\bibfnamefont {A.~P.}\ \bibnamefont {Sutton}},\ }\bibfield  {title}
  {\bibinfo {title} {{Electron-energy-loss spectra and the structural stability
  of nickel oxide: An LSDA+U study}},\ }\href
  {https://doi.org/10.1103/PhysRevB.57.1505} {\bibfield  {journal} {\bibinfo
  {journal} {Phys. Rev. B}\ }\textbf {\bibinfo {volume} {57}},\ \bibinfo
  {pages} {1505} (\bibinfo {year} {1998})}\BibitemShut {NoStop}%
\bibitem [{\citenamefont {Bl{\"{o}}chl}\ \emph {et~al.}(1994)\citenamefont
  {Bl{\"{o}}chl}, \citenamefont {Jepsen},\ and\ \citenamefont
  {Andersen}}]{BlochlPRB94T}%
  \BibitemOpen
  \bibfield  {author} {\bibinfo {author} {\bibfnamefont {P.~E.}\ \bibnamefont
  {Bl{\"{o}}chl}}, \bibinfo {author} {\bibfnamefont {O.}~\bibnamefont
  {Jepsen}},\ and\ \bibinfo {author} {\bibfnamefont {O.~K.}\ \bibnamefont
  {Andersen}},\ }\bibfield  {title} {\bibinfo {title} {{Improved tetrahedron
  method for Brillouin-zone integrations}},\ }\href
  {https://doi.org/10.1103/PhysRevB.49.16223} {\bibfield  {journal} {\bibinfo
  {journal} {Phys. Rev. B}\ }\textbf {\bibinfo {volume} {49}},\ \bibinfo
  {pages} {16223} (\bibinfo {year} {1994})}\BibitemShut {NoStop}%
\bibitem [{\citenamefont {Glazer}(1972)}]{GlazerACB72}%
  \BibitemOpen
  \bibfield  {author} {\bibinfo {author} {\bibfnamefont {A.~M.}\ \bibnamefont
  {Glazer}},\ }\bibfield  {title} {\bibinfo {title} {{The classification of
  tilted octahedra in perovskites}},\ }\href
  {https://doi.org/10.1107/S0567740872007976} {\bibfield  {journal} {\bibinfo
  {journal} {Acta Cryst. B}\ }\textbf {\bibinfo {volume} {28}},\ \bibinfo
  {pages} {3384} (\bibinfo {year} {1972})}\BibitemShut {NoStop}%
\bibitem [{\citenamefont {Biswas}\ \emph {et~al.}(2014)\citenamefont {Biswas},
  \citenamefont {Kim},\ and\ \citenamefont {Jeong}}]{BiswasJAP14}%
  \BibitemOpen
  \bibfield  {author} {\bibinfo {author} {\bibfnamefont {A.}~\bibnamefont
  {Biswas}}, \bibinfo {author} {\bibfnamefont {K.-S.}\ \bibnamefont {Kim}},\
  and\ \bibinfo {author} {\bibfnamefont {Y.~H.}\ \bibnamefont {Jeong}},\
  }\bibfield  {title} {\bibinfo {title} {{Metal insulator transitions in
  perovskite SrIrO$_3$ thin films}},\ }\href
  {https://doi.org/10.1063/1.4903314} {\bibfield  {journal} {\bibinfo
  {journal} {J. Appl. Phys.}\ }\textbf {\bibinfo {volume} {116}},\ \bibinfo
  {pages} {213704} (\bibinfo {year} {2014})}\BibitemShut {NoStop}%
\bibitem [{\citenamefont {Gruenewald}\ \emph {et~al.}(2014)\citenamefont
  {Gruenewald}, \citenamefont {Nichols}, \citenamefont {Terzic}, \citenamefont
  {Cao}, \citenamefont {Brill},\ and\ \citenamefont {Seo}}]{GruenewaldJMR14}%
  \BibitemOpen
  \bibfield  {author} {\bibinfo {author} {\bibfnamefont {J.~H.}\ \bibnamefont
  {Gruenewald}}, \bibinfo {author} {\bibfnamefont {J.}~\bibnamefont {Nichols}},
  \bibinfo {author} {\bibfnamefont {J.}~\bibnamefont {Terzic}}, \bibinfo
  {author} {\bibfnamefont {G.}~\bibnamefont {Cao}}, \bibinfo {author}
  {\bibfnamefont {J.~W.}\ \bibnamefont {Brill}},\ and\ \bibinfo {author}
  {\bibfnamefont {S.~S.}\ \bibnamefont {Seo}},\ }\bibfield  {title} {\bibinfo
  {title} {{Compressive strain-induced metal-insulator transition in
  orthorhombic SrIrO$_3$ thin films}},\ }\href
  {https://doi.org/10.1557/jmr.2014.288} {\bibfield  {journal} {\bibinfo
  {journal} {Journal of Materials Research}\ }\textbf {\bibinfo {volume}
  {29}},\ \bibinfo {pages} {2491} (\bibinfo {year} {2014})}\BibitemShut
  {NoStop}%
\bibitem [{\citenamefont {Fan}\ and\ \citenamefont {Yunoki}(2015)}]{FanJPCS15}%
  \BibitemOpen
  \bibfield  {author} {\bibinfo {author} {\bibfnamefont {W.}~\bibnamefont
  {Fan}}\ and\ \bibinfo {author} {\bibfnamefont {S.}~\bibnamefont {Yunoki}},\
  }\bibfield  {title} {\bibinfo {title} {{Electronic and Magnetic Structure
  under Lattice Distortion in SrIrO$_3$/SrTiO$_3$ Superlattice: A
  First-Principles Study}},\ }\href
  {https://doi.org/10.1088/1742-6596/592/1/012139} {\bibfield  {journal}
  {\bibinfo  {journal} {Journal of Physics: Conference Series}\ }\textbf
  {\bibinfo {volume} {592}},\ \bibinfo {pages} {012139} (\bibinfo {year}
  {2015})}\BibitemShut {NoStop}%
\bibitem [{\citenamefont {Watanabe}\ \emph {et~al.}(2014)\citenamefont
  {Watanabe}, \citenamefont {Shirakawa},\ and\ \citenamefont
  {Yunoki}}]{WatanabePRB14}%
  \BibitemOpen
  \bibfield  {author} {\bibinfo {author} {\bibfnamefont {H.}~\bibnamefont
  {Watanabe}}, \bibinfo {author} {\bibfnamefont {T.}~\bibnamefont
  {Shirakawa}},\ and\ \bibinfo {author} {\bibfnamefont {S.}~\bibnamefont
  {Yunoki}},\ }\bibfield  {title} {\bibinfo {title} {{Theoretical study of
  insulating mechanism in multiorbital Hubbard models with a large spin-orbit
  coupling: Slater versus Mott scenario in Sr$_2$IrO$_4$}},\ }\href
  {https://doi.org/10.1103/PhysRevB.89.165115} {\bibfield  {journal} {\bibinfo
  {journal} {Phys. Rev. B}\ }\textbf {\bibinfo {volume} {89}},\ \bibinfo
  {pages} {165115} (\bibinfo {year} {2014})}\BibitemShut {NoStop}%
\bibitem [{\citenamefont {Sch{\"{u}}tz}\ \emph {et~al.}(2017)\citenamefont
  {Sch{\"{u}}tz}, \citenamefont {{Di Sante}}, \citenamefont {Dudy},
  \citenamefont {Gabel}, \citenamefont {St{\"{u}}binger}, \citenamefont {Kamp},
  \citenamefont {Huang}, \citenamefont {Capone}, \citenamefont {Husanu},
  \citenamefont {Strocov}, \citenamefont {Sangiovanni}, \citenamefont {Sing},\
  and\ \citenamefont {Claessen}}]{SchutzPRL17}%
  \BibitemOpen
  \bibfield  {author} {\bibinfo {author} {\bibfnamefont {P.}~\bibnamefont
  {Sch{\"{u}}tz}}, \bibinfo {author} {\bibfnamefont {D.}~\bibnamefont {{Di
  Sante}}}, \bibinfo {author} {\bibfnamefont {L.}~\bibnamefont {Dudy}},
  \bibinfo {author} {\bibfnamefont {J.}~\bibnamefont {Gabel}}, \bibinfo
  {author} {\bibfnamefont {M.}~\bibnamefont {St{\"{u}}binger}}, \bibinfo
  {author} {\bibfnamefont {M.}~\bibnamefont {Kamp}}, \bibinfo {author}
  {\bibfnamefont {Y.}~\bibnamefont {Huang}}, \bibinfo {author} {\bibfnamefont
  {M.}~\bibnamefont {Capone}}, \bibinfo {author} {\bibfnamefont {M.-A.}\
  \bibnamefont {Husanu}}, \bibinfo {author} {\bibfnamefont {V.~N.}\
  \bibnamefont {Strocov}}, \bibinfo {author} {\bibfnamefont {G.}~\bibnamefont
  {Sangiovanni}}, \bibinfo {author} {\bibfnamefont {M.}~\bibnamefont {Sing}},\
  and\ \bibinfo {author} {\bibfnamefont {R.}~\bibnamefont {Claessen}},\
  }\bibfield  {title} {\bibinfo {title} {{Dimensionality-Driven Metal-Insulator
  Transition in Spin-Orbit-Coupled SrIrO$_3$}},\ }\href
  {https://doi.org/10.1103/PhysRevLett.119.256404} {\bibfield  {journal}
  {\bibinfo  {journal} {Phys. Rev. Lett.}\ }\textbf {\bibinfo {volume} {119}},\
  \bibinfo {pages} {256404} (\bibinfo {year} {2017})}\BibitemShut {NoStop}%
\bibitem [{\citenamefont {Zhang}\ \emph {et~al.}(2014)\citenamefont {Zhang},
  \citenamefont {Chen}, \citenamefont {Zhang}, \citenamefont {Zhou},
  \citenamefont {Zhang}, \citenamefont {Gu}, \citenamefont {Yao},\ and\
  \citenamefont {Chen}}]{ZhangJPSJ14}%
  \BibitemOpen
  \bibfield  {author} {\bibinfo {author} {\bibfnamefont {L.}~\bibnamefont
  {Zhang}}, \bibinfo {author} {\bibfnamefont {Y.~B.}\ \bibnamefont {Chen}},
  \bibinfo {author} {\bibfnamefont {B.}~\bibnamefont {Zhang}}, \bibinfo
  {author} {\bibfnamefont {J.}~\bibnamefont {Zhou}}, \bibinfo {author}
  {\bibfnamefont {S.}~\bibnamefont {Zhang}}, \bibinfo {author} {\bibfnamefont
  {Z.}~\bibnamefont {Gu}}, \bibinfo {author} {\bibfnamefont {S.}~\bibnamefont
  {Yao}},\ and\ \bibinfo {author} {\bibfnamefont {Y.}~\bibnamefont {Chen}},\
  }\bibfield  {title} {\bibinfo {title} {{Sensitively Temperature-Dependent
  Spin-Orbit Coupling in SrIrO$_3$ Thin Films}},\ }\href
  {https://doi.org/10.7566/JPSJ.83.054707} {\bibfield  {journal} {\bibinfo
  {journal} {J. Phys. Soc. Jpn.}\ }\textbf {\bibinfo {volume} {83}},\ \bibinfo
  {pages} {054707} (\bibinfo {year} {2014})}\BibitemShut {NoStop}%
\bibitem [{Note1()}]{Note1}%
  \BibitemOpen
  \bibinfo {note} {We have used $\epsilon ^\protect \text {LAO}/\epsilon _0 =
  24$, $\epsilon _0$ being the permittivity of free space, $\varepsilon _g +
  \Delta = 0.35$~eV, $K^\protect \text {IF} = 0.5$~F$^{-1}$}\BibitemShut
  {NoStop}%
\bibitem [{\citenamefont {Krupin}\ \emph {et~al.}(2009)\citenamefont {Krupin},
  \citenamefont {Bihlmayer}, \citenamefont {D{\"{o}}brich}, \citenamefont
  {Prieto}, \citenamefont {Starke}, \citenamefont {Gorovikov}, \citenamefont
  {Bl{\"{u}}gel}, \citenamefont {Kevan},\ and\ \citenamefont
  {Kaindl}}]{KrupinNJP09}%
  \BibitemOpen
  \bibfield  {author} {\bibinfo {author} {\bibfnamefont {O.}~\bibnamefont
  {Krupin}}, \bibinfo {author} {\bibfnamefont {G.}~\bibnamefont {Bihlmayer}},
  \bibinfo {author} {\bibfnamefont {K.~M.}\ \bibnamefont {D{\"{o}}brich}},
  \bibinfo {author} {\bibfnamefont {J.~E.}\ \bibnamefont {Prieto}}, \bibinfo
  {author} {\bibfnamefont {K.}~\bibnamefont {Starke}}, \bibinfo {author}
  {\bibfnamefont {S.}~\bibnamefont {Gorovikov}}, \bibinfo {author}
  {\bibfnamefont {S.}~\bibnamefont {Bl{\"{u}}gel}}, \bibinfo {author}
  {\bibfnamefont {S.}~\bibnamefont {Kevan}},\ and\ \bibinfo {author}
  {\bibfnamefont {G.}~\bibnamefont {Kaindl}},\ }\bibfield  {title} {\bibinfo
  {title} {{Rashba effect at the surfaces of rare-earth metals and their
  monoxides}},\ }\href {https://doi.org/10.1088/1367-2630/11/1/013035}
  {\bibfield  {journal} {\bibinfo  {journal} {New Journal of Physics}\ }\textbf
  {\bibinfo {volume} {11}},\ \bibinfo {pages} {013035} (\bibinfo {year}
  {2009})}\BibitemShut {NoStop}%
\bibitem [{\citenamefont {Rashba}(1960)}]{RashbaSPSS60}%
  \BibitemOpen
  \bibfield  {author} {\bibinfo {author} {\bibfnamefont {E.~I.}\ \bibnamefont
  {Rashba}},\ }\bibfield  {title} {\bibinfo {title} {{Properties of
  semiconductors with an extremum loop. 1. Cyclotron and combinational
  resonance in a magnetic field perpendicular to the plane of the loop}},\
  }\href@noop {} {\bibfield  {journal} {\bibinfo  {journal} {Sov. Phys. Solid
  State}\ }\textbf {\bibinfo {volume} {2}},\ \bibinfo {pages} {1109} (\bibinfo
  {year} {1960})}\BibitemShut {NoStop}%
\bibitem [{\citenamefont {Dresselhaus}(1955)}]{DresselhausPR55}%
  \BibitemOpen
  \bibfield  {author} {\bibinfo {author} {\bibfnamefont {G.}~\bibnamefont
  {Dresselhaus}},\ }\bibfield  {title} {\bibinfo {title} {{Spin-Orbit Coupling
  Effects in Zinc Blende Structures}},\ }\href
  {https://doi.org/10.1103/PhysRev.100.580} {\bibfield  {journal} {\bibinfo
  {journal} {Phys. Rev.}\ }\textbf {\bibinfo {volume} {100}},\ \bibinfo {pages}
  {580} (\bibinfo {year} {1955})}\BibitemShut {NoStop}%
\bibitem [{\citenamefont {Manca}\ \emph {et~al.}(2018)\citenamefont {Manca},
  \citenamefont {Groenendijk}, \citenamefont {Pallecchi}, \citenamefont
  {Autieri}, \citenamefont {Tang}, \citenamefont {Telesio}, \citenamefont
  {Mattoni}, \citenamefont {McCollam}, \citenamefont {Picozzi},\ and\
  \citenamefont {Caviglia}}]{MancaPRB18}%
  \BibitemOpen
  \bibfield  {author} {\bibinfo {author} {\bibfnamefont {N.}~\bibnamefont
  {Manca}}, \bibinfo {author} {\bibfnamefont {D.~J.}\ \bibnamefont
  {Groenendijk}}, \bibinfo {author} {\bibfnamefont {I.}~\bibnamefont
  {Pallecchi}}, \bibinfo {author} {\bibfnamefont {C.}~\bibnamefont {Autieri}},
  \bibinfo {author} {\bibfnamefont {L.~M.~K.}\ \bibnamefont {Tang}}, \bibinfo
  {author} {\bibfnamefont {F.}~\bibnamefont {Telesio}}, \bibinfo {author}
  {\bibfnamefont {G.}~\bibnamefont {Mattoni}}, \bibinfo {author} {\bibfnamefont
  {A.}~\bibnamefont {McCollam}}, \bibinfo {author} {\bibfnamefont
  {S.}~\bibnamefont {Picozzi}},\ and\ \bibinfo {author} {\bibfnamefont {A.~D.}\
  \bibnamefont {Caviglia}},\ }\bibfield  {title} {\bibinfo {title} {{Balanced
  electron-hole transport in spin-orbit semimetal SrIrO$_3$
  heterostructures}},\ }\href {https://doi.org/10.1103/PhysRevB.97.081105}
  {\bibfield  {journal} {\bibinfo  {journal} {Phys. Rev. B}\ }\textbf {\bibinfo
  {volume} {97}},\ \bibinfo {pages} {081105} (\bibinfo {year}
  {2018})}\BibitemShut {NoStop}%
\bibitem [{\citenamefont {van Thiel}\ \emph {et~al.}(2020)\citenamefont {van
  Thiel}, \citenamefont {Fowlie}, \citenamefont {Autieri}, \citenamefont
  {Manca}, \citenamefont {{\v{S}}i{\v{s}}kins}, \citenamefont {Afanasiev},
  \citenamefont {Gariglio},\ and\ \citenamefont {Caviglia}}]{VanThielACSML20}%
  \BibitemOpen
  \bibfield  {author} {\bibinfo {author} {\bibfnamefont {T.~C.}\ \bibnamefont
  {van Thiel}}, \bibinfo {author} {\bibfnamefont {J.}~\bibnamefont {Fowlie}},
  \bibinfo {author} {\bibfnamefont {C.}~\bibnamefont {Autieri}}, \bibinfo
  {author} {\bibfnamefont {N.}~\bibnamefont {Manca}}, \bibinfo {author}
  {\bibfnamefont {M.}~\bibnamefont {{\v{S}}i{\v{s}}kins}}, \bibinfo {author}
  {\bibfnamefont {D.}~\bibnamefont {Afanasiev}}, \bibinfo {author}
  {\bibfnamefont {S.}~\bibnamefont {Gariglio}},\ and\ \bibinfo {author}
  {\bibfnamefont {A.~D.}\ \bibnamefont {Caviglia}},\ }\bibfield  {title}
  {\bibinfo {title} {{Coupling Lattice Instabilities Across the Interface in
  Ultrathin Oxide Heterostructures}},\ }\href
  {https://doi.org/10.1021/acsmaterialslett.9b00540} {\bibfield  {journal}
  {\bibinfo  {journal} {ACS Materials Letters}\ }\textbf {\bibinfo {volume}
  {2}},\ \bibinfo {pages} {389} (\bibinfo {year} {2020})}\BibitemShut {NoStop}%
\bibitem [{\citenamefont {Zhong}\ \emph {et~al.}(2013)\citenamefont {Zhong},
  \citenamefont {T{\'{o}}th},\ and\ \citenamefont {Held}}]{ZhongPRB13}%
  \BibitemOpen
  \bibfield  {author} {\bibinfo {author} {\bibfnamefont {Z.}~\bibnamefont
  {Zhong}}, \bibinfo {author} {\bibfnamefont {A.}~\bibnamefont {T{\'{o}}th}},\
  and\ \bibinfo {author} {\bibfnamefont {K.}~\bibnamefont {Held}},\ }\bibfield
  {title} {\bibinfo {title} {{Theory of spin-orbit coupling at
  LaAlO$_3$/SrTiO$_3$ interfaces and SrTiO$_3$ surfaces}},\ }\href
  {https://doi.org/10.1103/PhysRevB.87.161102} {\bibfield  {journal} {\bibinfo
  {journal} {Phys. Rev. B}\ }\textbf {\bibinfo {volume} {87}},\ \bibinfo
  {pages} {161102} (\bibinfo {year} {2013})}\BibitemShut {NoStop}%
\end{thebibliography}
\end{document}